\documentclass[aps,prb,twocolumn,groupedaddress]{revtex4}

\usepackage{amsmath}
\usepackage{amsfonts}
\usepackage{siunitx}
\usepackage{xcolor}
\usepackage{graphicx}
\usepackage{bm}
\usepackage[caption=false]{subfig}
\usepackage[export]{adjustbox}

\makeatletter
\def\input@path{{Images/}}
\makeatother

\newcommand{\ve}{\varepsilon}

\begin{document}

\title{Patterned bilayer graphene as a tunable, strongly correlated system}

\author{Z. E. Krix}
\email[]{z.krix@unsw.edu.au}
\author{O. P. Sushkov}
\affiliation{School of Physics, University of New South Wales, Sydney 2052}
\affiliation{Australian Research Council Centre of Excellence in Low-Energy Electronics Technologies, University of New South Wales, Sydney 2052, Australia}

\date{\today}

\begin{abstract}
    Recent observations of superconductivity in Moire graphene \cite{cao_unconventional_2018} have lead to an intense interest in that system, with subsequent studies revealing a more complex phase diagram including correlated insulators and ferromagnetic phases. Here we propose an alternate system, electrostatically patterned bilayer graphene (PBG), in which a supermodulation is induced via metallic gates rather than the moire effect. We show that, by varying either the gap or the modulation strength, bilayer graphene can be tuned into the strongly correlated regime. Further calculations show that this is not possible in monolayer graphene. We present a general technique for addressing Coulomb screening of the periodic potential and demonstrate that this system is experimentally feasible.
\end{abstract}

\maketitle

\section{Introduction}

Superconductivity in twisted bilayer graphene \cite{cao_unconventional_2018} occurs at a twist angle which turns the lowest lying energy states into a flat band \cite{bistritzer_moire_2011}. More generally, strongly correlated phases due to flat band physics arise in a broad range of materials. Observations of superconductivity \cite{cao_unconventional_2018, lu_superconductors_2019, yankowitz_tuning_2019, chen_signatures_2019, park_tunable_2021}, correlated insulators \cite{cao_correlated_2018, lu_superconductors_2019, yankowitz_tuning_2019, kerelsky_maximized_2019, chen_signatures_2019, chen_evidence_2019}, ferromagnetism \cite{sharpe_emergent_2019, serlin_intrinsic_2020} and nematic order \cite{kerelsky_maximized_2019, jiang_charge_2019, rubio-verdu_moire_2021} have been reported across the family of twisted graphene systems. This includes twisted bilayer graphene \cite{cao_unconventional_2018}, twisted trilayer graphene \cite{park_tunable_2021}, and twisted double bilayer graphene \cite{rubio-verdu_moire_2021}. Flat bands also arise in twisted TMDCs \cite{zhang_flat_2020} and kagome systems which exhibit superconductivity, ferromagnetism, and charge density waves \cite{lin_flatbands_2018, ortiz_cs_2020, ortiz_superconductivity_2021, chen_double_2021, zhu_double-dome_2021, jiang_unconventional_2021}.

Given the high level of interest in strongly correlated phases arising from flat band systems, particularly twisted bilayer graphene, the present work proposes an alternative graphene-based system which is fully tunable and contains a well-defined, isolated flat band. The system we consider is a graphene bilayer with no twist angle and a patterned electrostatic gate a vertical distance, $z$, from the bilayer. For brevity we refer to this system as patterned bilayer graphene (PBG). The guiding idea is to restructure the bare energy bands of bilayer graphene via periodic electrostatic gating rather than with a twist-induced Moire superlattice. Conceptually, this is a continuation of our previous work on semiconductor artificial crystals \cite{krix_correlated_2022}, which are less efficient than PBG at generating a strong modulation. A major advantage of this approach is that it bypasses the issue of twist-angle disorder \cite{uri_mapping_2020} (i.e. long-range spatial variation of the twist angle). The Moire flat band occurs at a precise value of twist angle ($\theta \approx \SI{1.1}{\degree}$) and a modest amount of twist-disorder ($\lesssim 10 \ \%$) can destroy this band \cite{wilson_disorder_2020}. We demonstrate that PBG has no equivalent fine-tuning or disorder problem.

The central advantage of our system is its controllability. A designed superlattice potential induced by patterned electrostatic gating can have any desired lattice symmetry (e.g. square, triangular, honeycomb\cite{polini_artificial_2013}, Lieb, or kagome) and lattice constants as small as \SI{40}{\nano \metre} \cite{huber_gate-tunable_2020}. It is also possible to tune both the strength of the supermodulation and the particle density independently \cite{huber_gate-tunable_2020, wang_two-dimensional_2020}. In contrast, Moire graphene superlattices have a triangular symmetry which is fixed by the crystal structure of graphene. The superlattice constant, $a \approx \SI{13}{\nano \metre}$, is also fixed by the flat band condition, $\theta \approx \SI{1.1}{\degree}$ and tuning the superlattice strength is only possible by applying hydrostatic pressure \cite{yankowitz_tuning_2019}. Some prior works have focused on patterning monolayer graphene, either by etching holes directly into the graphene sheet \cite{furst_electronic_2009} or by patterned electrostatic gating \cite{huber_gate-tunable_2020, huber_brown-zak_2021}. Ref. \cite{furst_electronic_2009} demonstrates, theoretically, that patterning introduces an energy gap in the graphene dispersion while Refs. \cite{huber_gate-tunable_2020, huber_brown-zak_2021} measure magnetotransport properties of a real device and show that the result of patterning is essentially a correction to single particle physics. There is not, however, the possibility for generating an isolated flat band or strongly correlated phases in these monolayer graphene systems.

Our results are derived from band structure calculations in a continuum, bilayer graphene model with imposed superlattice potential. We find that bilayer graphene can be driven into the Mott regime by application of a suffciently strong band gap and potential modulation. This occurs because a flat band develops in the lowest-energy band of the PBG dispersion. By detuning either the band gap or potential modulation the system can be tuned out of the Mott regime while keeping the total electron density fixed. Within the flat band it is possible to mimic the dispersion of many different two-dimensional lattices including square, triangular, kagome, and Lieb, by varying the symmetry of the patterned gate. We show, using an analogous calculation, that it is not possible to generate a flat band in monolayer graphene. Lastly, we study electron-electron screening in bilayer graphene. Current techniques are not able to address a system with both an unbounded dispersion and a strong potential modulation; we develop a general technique to address Coulomb screening in this limit. The technique we develop is general and could also be applied to, for example, the problem of impurity screening in bilayer graphene. We show that screening of the periodic potential is strong but can be overcome by experimentally realistic gate voltages. Our results show that patterned bilayer graphene is an experimentally viable way to engineer an isolated flat with almost complete control over the underlying effective Hubbard model.

\begin{figure}[t]
\caption{Sketch of the two Brillouin zones. The larger is that of the underlying bilayer graphene system and the smaller is that of the artificial crystal. Since we use an expansion about the $K$ points of the bilayer graphene BZ, the artificial BZ is centered at a $K$ point. The inset shows the artificial crystal in real space.}
\centering
\includegraphics[width=0.32\textwidth]{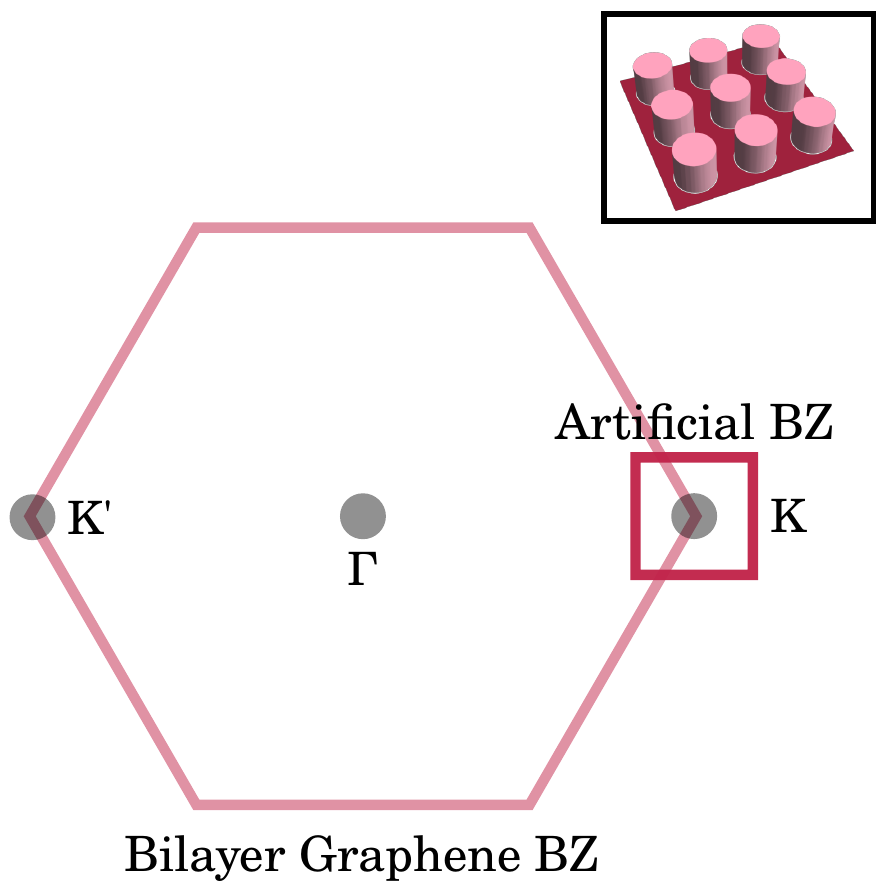}
\label{fig:CBLG_BZ}
\end{figure}

\section{Theoretical Techniques}\label{sec:CBLG_theoreticalTechniques}

Our starting point is a plain bilayer graphene sheet with external, perpendicular electric field, $\bm{E}$, inducing an energy gap, $\Delta$. The relationship between field and gap is roughly $|\bm{E}| \propto \Delta$, where the constant of proportionality is such that a displacement field of $\SI{1}{\volt \per \nano \meter}$ leads to a gap $\Delta = \SI{100}{\milli \electronvolt}$.\cite{zhang_direct_2009} We find that the value of the gap is important but does not need to be finely tuned, we discuss conditions on $\Delta$ below. The low-energy effective Hamiltonian for a single valley of bilayer graphene is \cite{mccann_electronic_2013}

\begin{align}\label{equ:CBLG_blgBareHam}
    H_{BLG} =
    \begin{bmatrix}
        \Delta / 2 & v p_{-}    & 0            & \gamma  \\
        v p_{+}    & \Delta / 2 & 0            & 0       \\
        0          & 0          & - \Delta / 2 & v p_{-} \\
        \gamma     & 0          & v p_{+}      & - \Delta / 2
    \end{bmatrix}
\end{align}

Where $v \approx \SI{1e6}{\meter \per \second}$ is the Fermi velocity of monolayer graphene, $\gamma \approx \SI{0.38}{\electronvolt}$ is the coupling between graphene layers, and the operator $p_{\pm}$ is defined by $p_{\pm} = p_{x} \pm i p_{y}$. These values are taken from Ref. \cite{mccann_electronic_2013}. The Hamiltonian is composed of $2\times2$ blocks. Each diagonal block is the Hamiltonian of a single graphene layer, and each layer has a different energy shift $\pm \Delta / 2$ depending on its position in the external field. The off-diagonal blocks, which couple the two layers, arise from the simplest kind of interlayer hopping, between two carbon atoms which are vertically aligned: $\gamma$ is the matrix element for this hopping. One can also include terms in the Hamiltonian which describe longer-range interlayer hopping (these are denoted by $\gamma_{3}$ and $\gamma_{4}$ in Ref. \cite{mccann_electronic_2013}). As discussed in Ref. \cite{mccann_electronic_2013} they contribute to trigonal warping and particle-antiparticle asymmetry of the band dispersion. These additional terms are secondary to the major terms, $v$ and $\gamma$, and, for the sake of physical transparency, we neglect them here. In the limit $|\ve| \ll \gamma$, the two low-energy bands which arise from Eqn. \ref{equ:CBLG_blgBareHam} are roughly quadratic: $\ve(p) = \pm \sqrt{ (p^{2} / 2 m^{*})^{2} + \Delta^{2} / 4 }$, where the effective mass is $m^{*} = \gamma / 2 v^{2} \approx 0.03$.

Over the top of this Hamiltonian we wish to introduce a spatially modulated electrostatic potential, $U(\bm{r})$, due to the patterned gating. Suppose, first, that the periodic potential defined \emph{at the gate} is given by

\begin{align*}
    U_{\text{gate}}(\bm{r}) =
    W
    \sum_{\bm{G}}
    e^{i \bm{G} \cdot \bm{r}}
    U_{\bm{G}}
\end{align*}

Where the vectors, $\bm{G}$, are the reciprocal lattice vectors of the artificial superlattice and $W$ is a parameter taking dimensions of energy which controls the strength of the superlattice. In our calculation, the dimensionless parameters, $U_{\bm{G}}$, define a muffin-tin shaped, periodic potential with square-lattice symmetry (see the inset to Fig. \ref{fig:CBLG_BZ}). In this case $W$ is equal to the total variation in potential energy from minimum to maximum, the ``height'' of the muffin-tin. For concreteness we study $W > 0$, which corresponds to an array of anti-dots. The opposite limit $W < 0$ is very similar and we discuss this below. The potential, $U(\bm{r})$, at the plane of the bilayer graphene sheet is then

\begin{align}\label{equ:CBLG_potentialAtBLG}
    U(\bm{r}) =
    W
    \sum_{\bm{G}}
    e^{i \bm{G} \cdot \bm{r}}
    e^{- G z}
    U_{\bm{G}}
\end{align}

Where $z$ is the vertical distance between the patterned gate and the bilayer. This exponential suppression of the higher harmonics follows directly from Poisson's equation. For a square lattice, the first harmonic has $G = g = 2 \pi / a$, where $a$ is the superlattice period. We choose parameters $a = \SI{80}{\nano \meter}$ and $z = \SI{10}{\nano \meter}$ which are reasonable from an experimental standpoint. The supression of the second harmonic relative to the first is then $e^{- g z} \sim 0.5$ meaning that higher harmonics should not be neglected. For a smaller lattice constant, $a = \SI{40}{\nano \meter}$, $e^{-gz} \sim 0.2$ and so higher harmonics are slightly less significant. Given this periodic potential the total Hamiltonian becomes

\begin{align*}
    H_{PBG} =
    H_{BLG} + U(\bm{r})
\end{align*}

The matrix structure of $U(\bm{r})$ is trivial, it is a $4 \times 4$ identity matrix. Strictly speaking there should also be a spatially varying correction to the gap, $\Delta$, due to each graphene layer being a slightly different distance from the patterned gate. This correction, however, produces a small effect relative to the major contribution, $U(\bm{r})$, and can be neglected. Note that the applied, constant field (which induces $\Delta$) will give a similar, diagonal contribution to the Hamiltonian. Since that contribution is spatially invariant, it simply shifts all energy levels by the same amount, and can be left out of the Hamiltonian.

Since the artificial superlattice period, $a$, is larger than the period of the graphene lattice and since we are focusing on a single valley, our approach amounts to defining a smaller, artificial Brillouin zone (either square or hexagonal) around one of the vertices of the original bilayer graphene Brillouin zone. This is sketched in figure \ref{fig:CBLG_BZ}.

\begin{figure}[b]
\caption{Evolution of dispersion from $W = \SI{0}{\milli \electronvolt}$ to $W = \SI{30}{\milli \electronvolt}$. In each panel $a = \SI{80}{\nano \meter}$, $\Delta = \SI{15}{\milli \electronvolt}$, $R = \SI{15}{\nano \meter}$ and $z = \SI{10}{\nano \meter}$. At $W = \SI{0}{\milli \electronvolt}$ we have the bare BLG dispersion band-folded into the artificial Brillouin zone. By $W = \SI{30}{\milli \electronvolt}$ a single, flat band (width $\approx \SI{0.5}{\milli \electronvolt}$) has separated from the rest of the hole bands. Points around the artificial Brillouin zone are defined in the inset.}
\centering
\includegraphics[width=0.45\textwidth]{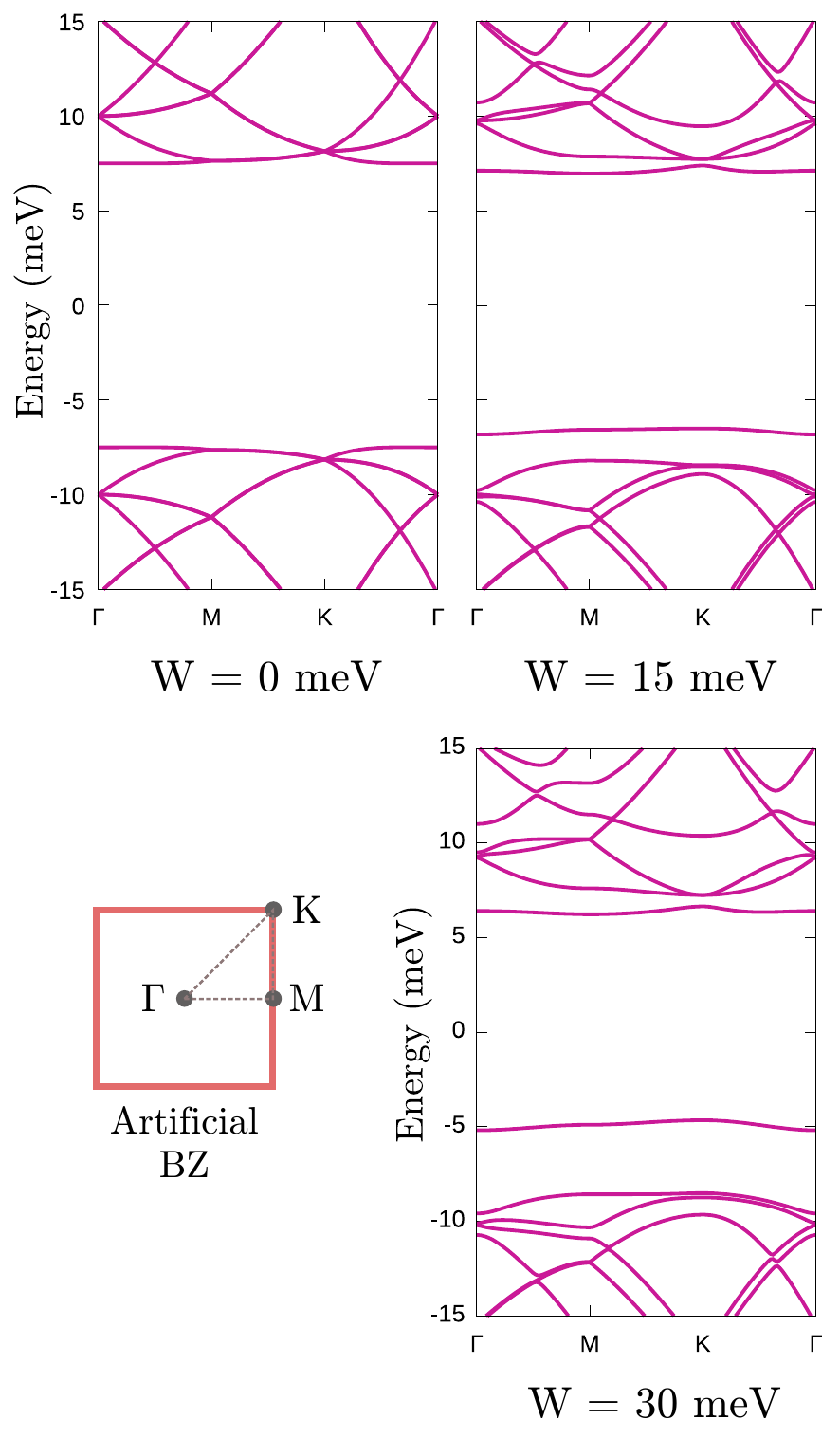}
\label{fig:CBLG_dispEvo}
\end{figure}

The energy levels of the Hamiltonian, $H_{PBG}$, can be obtained exactly by numerical diagonalisation. We must first compute the matrix elements of $H_{PBG}$ in a particular basis. For simplicity, we choose basis vectors, $| \bm{k} + \bm{G}, i \rangle$, defined by

\begin{align*}
    | \bm{k} + \bm{G}, i \rangle =
    \frac{1}{\sqrt{A}}
    e^{i (\bm{k} + \bm{G}) \cdot \bm{r}}
    | i \rangle \\  \quad
    ( | i \rangle )_{j} = \delta_{ij}, \quad
    i, \ j = 1, \cdots, 4
\end{align*}

Here $| i \rangle$ is a 4-tuple, $\bm{G}$ is a reciprocal lattice vector of the superlattice and $\bm{k}$ is the quasi-momentum, which sits somewhere within the \emph{artificial} Brillouin zone. The normalisation factor, $A$, is just the total area of the sample. For example, at $i = 1$, we have

\begin{align*}
    | 1 \rangle =
    \begin{bmatrix}
        1 \\ 0 \\ 0 \\ 0
    \end{bmatrix}
\end{align*}

The matrix elements of $H_{PBG}$ are then

\begin{align*}
    \langle \bm{k} + \bm{G}, i |
    H_{PBG}
    | \bm{k} + \bm{G}', j \rangle
    = &
    \langle i | H_{BLG} ( \bm{k} + \bm{G} ) | j \rangle
    \delta_{\bm{G}, \bm{G}'} \\
    + &
    \delta_{i,j}
    W U_{\bm{G} - \bm{G}'}
    e^{ - |\bm{G} - \bm{G}'| z }
\end{align*}

In $H_{BLG}( \bm{k} )$ the operator, $p_{\pm}$, is replaced by the complex number, $k_{\pm} = k_{x} \pm i k_{y}$. The matrix elements of the potential are given by,

\begin{align*}
    U_{\bm{G}} =
    2 \pi \frac{R^{2}}{A_{cell}}
    \frac{ J_{1}( | \bm{G} | R)  }{ | \bm{G} | R }
\end{align*}

Where $R$ is the radius of a single anti-dot, taken to be $R = \SI{15}{nm}$ in our calculations, and $A_{cell}$ is the unit cell area of the artificial crystal; $J_{1}$ is a Bessel function of the first kind. The only differences between square an triangular lattices are the choice of reciprocal lattice vectors, $\bm{G}$, and the size of $A_{cell}$ in the above equation. To diagonalise this Hamiltonian numerically we must make the set of basis vectors finite by choosing a maximal $\bm{G}$ vector. In practice, we simply increase the size of the basis until the energy levels or eigenvectors converge. The result of this procedure is a set of energy levels, $\ve_{n}( \bm{k} )$, and eigenvectors, $\psi_{n,\bm{k}}(\bm{r})$, each being a function of the quasi-momentum, $\bm{k}$, of the artificial superlattice. The band index can take values $n = \pm 1, \pm 2, \cdots$ with the charge neutrality point occurring between $n = -1$ and $n = +1$. Each band, $n$, is degenerate across valleys and spins and contains a number of particles, $n_{0} = 4 / A_{cell}$, corresponding to complete filling of one artificial Brillouin zone (accounting for spin and valley degeneracy). At $a = \SI{80}{\nano \meter}$ we have $n_{0} = \SI{6.25e10}{\per \square \centi \meter}$, while at $a = \SI{40}{\nano \meter}$ we have $n_{0} = \SI{2.5e11}{\per \square \centi \meter}$.

The parameter $W$ has been defined at the patterned gate. This allows us to fairly compare the strength of a square lattice with that of a kagome lattice (for example). Still, it is useful to have a general idea of the amplitude of the potential at the 2DEG. Looking at the first harmonic only, this quantity is

\begin{align*}
    W_{2DEG} =
    W
    2 \pi \frac{R^{2}}{A_{cell}}
    \frac{ J_{1}( | \bm{G} | R)  }{ | \bm{G} | R }
    e^{-| \bm{G} | z}
\end{align*}

As a rough approximation, keeping only first harmonics, the potential at the 2DEG is $2 W_{2DEG} ( \cos(gx) + \cos(gy) )$. For the parameters we are using the conversion is $W_{2DEG} = 0.042 W$. At $W = \SI{30}{\milli \electronvolt}$ we have $W_{2DEG} = \SI{1.26}{\milli \electronvolt}$. If the total variation in potential energy is $\SI{30}{\milli \electronvolt}$ at the gate, then the equivalent quantity at the 2DEG is around $8W_{2DEG} = \SI{10}{\milli \electronvolt}$.

\section{Generating a flat band}

We will first demonstrate how the bandstructure evolves as $W$ is tuned, keeping the band gap fixed at $\Delta = \SI{15}{\milli \electronvolt}$. Our results are summarised in Fig. \ref{fig:CBLG_dispEvo}. In the first panel of Fig. \ref{fig:CBLG_dispEvo} $W = 0$ and the dispersion is just the result of band-folding the bare BLG dispersion into the artificial, square Brillouin zone: at $\Gamma$ the electron band starts at $+\SI{7.5}{\milli \electronvolt}$ and the hole band starts at $-\SI{7.5}{\milli \electronvolt}$. We find that as the potential is turned on the first hole band ($n = -1$) is separated from the remaining hole bands ($n = -2, \ -3, \ \cdots$) and develops a total band width $\delta_{b} = \SI{0.53}{\milli \electronvolt}$, which does not change as $W$ is increased beyond $\SI{30}{\milli \electronvolt}$. The existence of this flat band is the main focus of our work. As can be seen, the full dispersion is electron-hole asymmetric; we address this point later in the section.

\begin{figure}[t]
    \centering
\caption{Zoom-in of the $n = -1$ band in Fig. \ref{fig:CBLG_dispEvo}. Each curve corresponds to a different value of $W$, with $W = \SI{0}{\milli \electronvolt}$ for the lowest curve and $W = \SI{30}{\milli \electronvolt}$ for the upper most curve.}
\label{fig:CBLG_dispEvoFlatBand}
\includegraphics[width=0.4\textwidth]{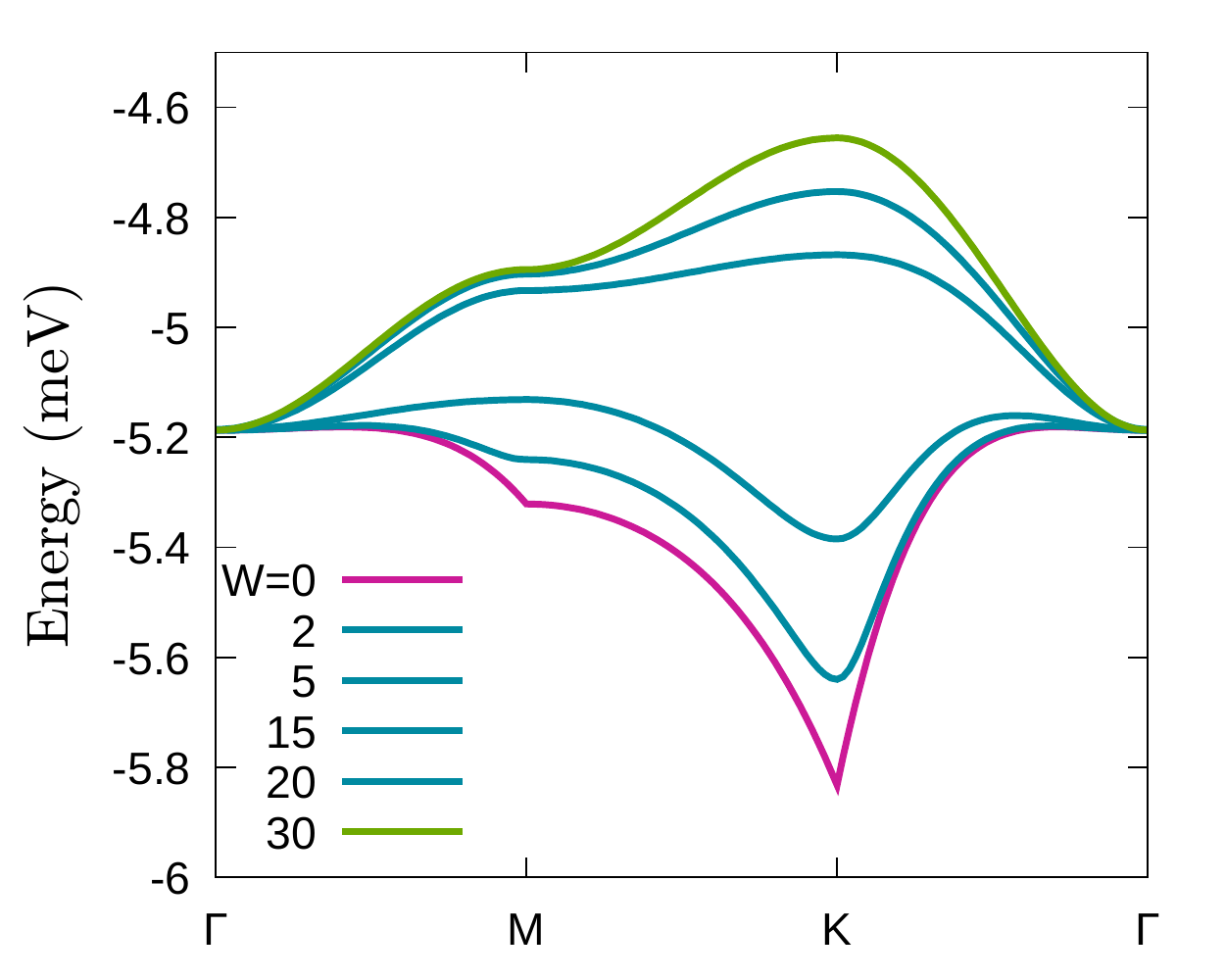}
\end{figure}

To study the flat band more closely, we plot it individually in Fig. \ref{fig:CBLG_dispEvoFlatBand} for a range of $W$ values. We find that the effect of the potential is not only to separate this band, but to transform its dispersion into that of a tight-binding model on a square lattice with nearest-neighbour hopping (green band in Fig. \ref{fig:CBLG_dispEvoFlatBand}). This dispersion is $2 t (\cos(a k_{x}) + \cos(a k_{y}))$ where $t$ is the matrix element for nearest-neighbour hopping. We can thus map the physics of the flat band to an effective tight-binding model by equating the total band width, $\delta_{b}$, with $8 t$: we find that the effective hopping parameter is $t = \SI{0.066}{\milli \electronvolt}$. The number of states within this band is $n_{0} = 4 / A_{cell} = \SI{6.25e10}{\per \square \centi \meter}$, with the factor $4$ arising from spin and valley degeneracy.

The emergence of a flat band is related to the energy gap. A plain bilayer graphene sheet has two, roughly quadratic bands with opposite curvature and the effect of an out-of-plane electric field is not only to open a gap between these bands, but to flatten the bottom of each band as well. The idea, then, is to first flatten the bottom of the band using a constant electric field and then to separate a flat band by imposing a periodic modulation. The artificial Brillouin zone ``cuts out'' the flat part of the hole band. We thus require that the first Brillouin zone of the artificial lattice fit within the flattened part of the bare BLG dispersion. To generate a flat band using a smaller superlattice period -- that is, with a larger Brillouin zone -- we must increase the size of the flat region in the bare BLG dispersion by increasing the gap. The presence of the gap makes it easier for an external potential to localise electrons, and also allows for separation between electron and hole bands so that the resulting flat band is isolated.

This suggests a criterion for the minimum band gap $\Delta$ which can give a flat band. If we approximate the bare dispersion by $\ve(k) = \sqrt{(p^{2}/2 m^{*})^{2} + \Delta^{2}/4}$ then the ``flat region'' of the dispersion is roughly defined by $p^{2} / m^{*} = \Delta$. If we want the boundary of this region to coincide with the Brillouin zone boundary then the condition on the band gap is $\Delta \sim (\pi / a)^{2} / m^{*} = \SI{3.5}{\milli \electronvolt} $.

\begin{figure}[t]
\caption{Evolution of dispersion from $\Delta = \SI{0}{\milli \electronvolt}$ to $\Delta = \SI{5}{\milli \electronvolt}$. In each panel $a = \SI{80}{\nano \meter}$, $W = \SI{30}{\milli \electronvolt}$, $R = \SI{15}{\nano \meter}$ and $z = \SI{10}{\nano \meter}$. The flat band becomes isolated by $\Delta = \SI{5}{\milli \electronvolt}$, but the dispersion $2 t ( \cos(k_{x} a) + \cos(k_{y} a) )$ does not develop until $\Delta = \SI{15}{\milli \electronvolt}$ (Fig. \ref{fig:CBLG_dispEvo}).}
\centering
\includegraphics[width=0.45\textwidth]{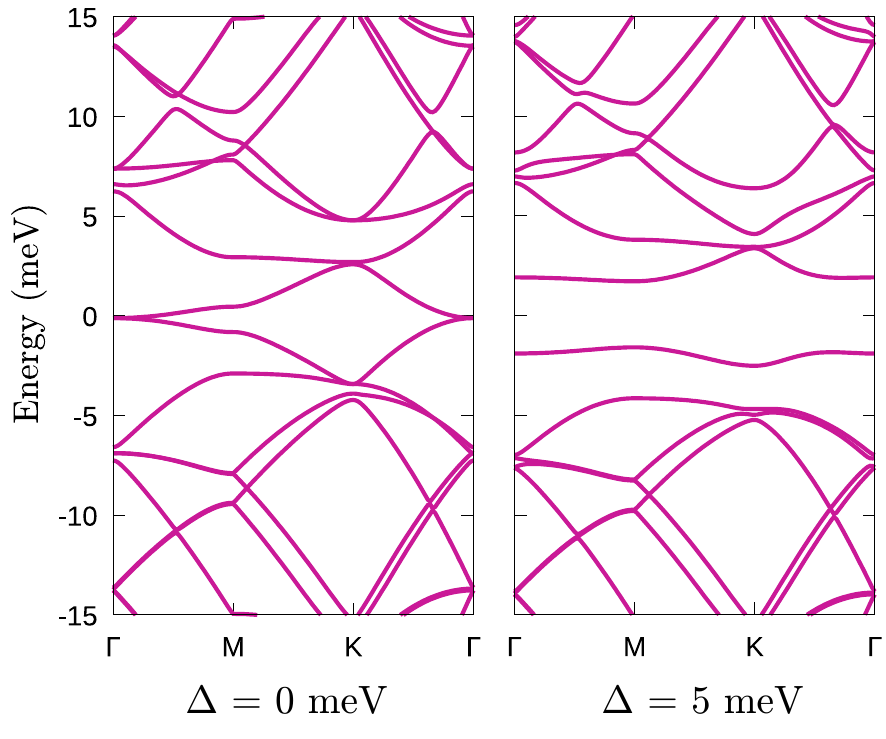}
\label{fig:CBLG_dispGapEvo}
\end{figure}

To demonstrate the importance of the band gap we compute the evolution of the bandstructure as $\Delta$ is tuned, keeping $W = \SI{30}{\milli \electronvolt}$ fixed. The results are plotted in Fig. \ref{fig:CBLG_dispGapEvo}. At $\Delta = 0$ the electron and hole bands touch and there is no isolated flat band. At $\Delta = \SI{5}{\milli \electronvolt}$ the $n = -1$ band becomes isolated. And finally, at $\Delta = \SI{15}{\milli \electronvolt}$ (Fig. \ref{fig:CBLG_dispEvo}), the dispersion of the $n = -1$ band matches that of the square-lattice tight binding model: $2 t ( \cos(k_{x} a) + \cos(k_{y} a) )$. Bilayer graphene can be tuned in and out of the flat band regime in two ways. First, one can fix the band gap and increase the potential modulation from zero. Second, one can fix the potential modulation and increase the gap.

The shape of the flat band dispersion follows from shape of the potential energy: a square array of hat functions. We have defined the potential such that electrons in the upper BLG band ($n > 0$) minimise their energy at minimal values of $U(\bm{r})$. Electrons in the lower BLG band thus minimise their energy by seeking the potential maxima, and can be localised to the square lattice by making the imposed potential sufficiently strong. We can investigate localisation by computing the particle density due to the sum of all electrons within the first hole band. That is, we can compute

\begin{align*}
    \rho_{n}(\bm{r}) =
    \sum_{\bm{k}, \sigma} |\psi_{n, \bm{k}, \sigma}(\bm{r})|^{2}
\end{align*}

\begin{figure}[t]
\caption{particle density of electrons in the flat band for different values $W$ at $\Delta = \SI{15}{\milli \electronvolt}$. (a) Map of particle density at $W = \SI{30}{\milli \electronvolt}$. (b) Cuts of particle density along $y = 0$ for $W = \SI{2}{\milli \electronvolt}$ to $W = \SI{30}{\milli \electronvolt}$.}
\centering
\includegraphics[width=0.37\textwidth]{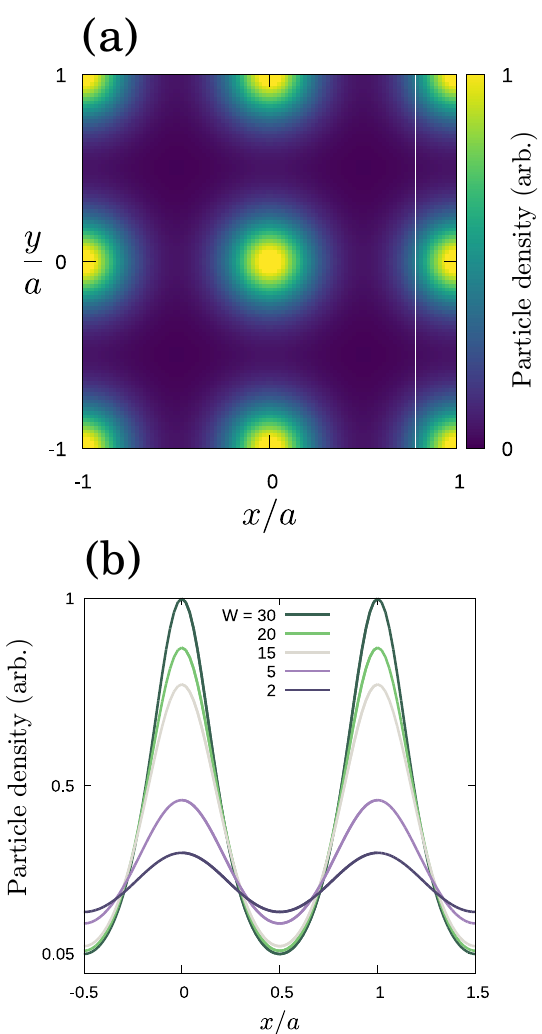}
\label{fig:CBLG_chargeDensityCutMap}
\end{figure}

Where the band index is $n = -1$. Our results are plotted in Fig. \ref{fig:CBLG_chargeDensityCutMap}. Panel (a) shows a map of the particle density for $W = \SI{30}{\milli \electronvolt}$. Here, the density is peaked at the anti-dot sites and has square-lattice symmetry, electrons are well localised to each site. To quantify this we present some cuts of the particle density along the nearest neighbour line between two anti-dot sites for different values of $W$ (Fig. \ref{fig:CBLG_chargeDensityCutMap}b). We find that as $W$ increases the particle density at the mid point between two sites -- a measure of the wavefunction overlap -- decreases until, at $W = \SI{30}{\milli \electronvolt}$, it is $5\%$ its maximum value. This strong localisation to a square lattice is the reason the dispersion reproduces that of a tight-binding model with small $t$ ($t = \SI{0.066}{\milli \electronvolt}$); it is enabled by the presence of a finite energy gap, $\Delta$. The localisation does not increase indefinitely as you increase $W$: once the electron is squeezed into an area the size of the anti-dot, further increase of $W$ does not increase the degree of localisation. Thus, the width of the particle density peak in Fig. \ref{fig:CBLG_chargeDensityCutMap}b is around $0.4 a \approx \SI{30}{\nano \meter} = 2 R$. Because localisation is limited to the size of the anti-dot, the effective hopping parameter, $t$, does not change any further.

Note that this explains why the dispersion in Fig. \ref{fig:CBLG_dispEvo} is particle-hole assymetric. Our potential (Fig. \ref{fig:CBLG_BZ}) has a series of well-defined maxima, which can localise particles in the hole band. The minima of the potential, which can localise particles in the electron band, arise from the Poisson factor, $e^{- G z}$, in Eqn. \ref{equ:CBLG_potentialAtBLG} and are thus weak. A larger value of $W$ is required to localise particles in the electron band, compared to particles in the hole band. We thus find that a flat dispersion develops in the hole bands before it develops in the electron bands, as $W$ is turned on. We could alternatively create a flat dispersion in the electron bands by imposing an array of dots rather than anti-dots: the transformation $W \mapsto - W$ flips all of the energy levels like $\ve \mapsto - \ve$.

\section{Mapping to Hubbard model}

The dispersion and the particle density of the flat band suggest a mapping to a tight-binding model on a square lattice. We have already shown that the band width of the flat band can be used to define an effective hopping parameter for the artificial crystal $t = \SI{0.066}{\milli \electronvolt}$. Given the particle density in Fig. \ref{fig:CBLG_chargeDensityCutMap} we can also define an effective, on-site Coulomb repulsion energy, $U_{H}$. We can take the form of the particle density around a single bright dot in Fig. \ref{fig:CBLG_chargeDensityCutMap}a and treat it as the wave function for a single electron, $|\psi(\bm{r})|^{2}$. This has to be normalised so that there is one electron per bright dot

\begin{align*}
    \int_{Cell} d^{2} \bm{r}
    |\psi(\bm{r})|^{2}
    = 1
\end{align*}

Having extracted the function $|\psi(\bm{r})|^{2}$ from figure \ref{fig:CBLG_chargeDensityCutMap} we can compute the effective Hubbard energy

\begin{align*}
    U_{H} =
    \frac{e^{2}}{\ve}
    \int d^{2} \bm{r}' d^{2} \bm{r}
    \frac{
        |\psi(\bm{r}')|^{2}
        |\psi(\bm{r})|^{2}
    }{|\bm{r}' - \bm{r}|}
\end{align*}

Where $\ve$ is roughly the dielectric constant of the encapsulating medium, which for the purpose of estimating $U_{H}$ we take to be hBN so that $\ve = 4$. We find that, at $W = \SI{30}{\milli \electronvolt}$, the particle density around a single bright dot is fit by a Gaussian $|\psi(\bm{r})|^{2} \propto e^{-r^{2}/\sigma}$ with $\sigma \approx (0.25 a)^{2} = 0.063 a^{2}$. After the appropriate normalisation this gives a Hubbard energy $U_{H} \approx \SI{10}{\milli \electronvolt}$. We thus have a dramatic ratio $U_{H} / t \approx 150$. Our estimate of $t$ is essentially exact, however the estimate of $U_{H}$ is an upper limit, its value will be reduced by higher order effects. The actual value of $U_{H} / t$ will then be less than this estimate.

One factor which reduces $U_{H}$ is screening by the patterned gate. Any charge in the graphene layers will induce an image charge in the patterned layer which screens the Coulomb interaction. The corresponding screening length is roughly the distance $z \approx \SI{10}{\nano \meter}$ between the gate and the bilayer graphene sheet. Accounting for this effect gives the following, corrected Coulomb energy

\begin{align*}
    \frac{e^{2}}{r} -
    \frac{e^{2}}{\sqrt{r^{2} + 4 z^{2}}}
\end{align*}

Repeating the estimate with this Coulomb energy gives the reduced value, $U_{H} \approx \SI{5}{\milli \electronvolt}$, so that the ratio with $t$ becomes $U_{H} / t \approx 75$. Additional effects will reduce this number further, perhaps by another $50\%$. What is clear from this rough estimate, however, is that the Mott regime, defined by $U_{H} / t \sim 7$, \cite{hubbard_electron_1964} is well within reach of experiments. In all of these estimates we have fixed the lattice constant at $a = \SI{80}{\nano \meter}$. Supposing that we could scale the system down in all directions by half, so that $a = \SI{40}{\nano \meter}$ and $R = \SI{7.5}{\nano \meter}$, we would expect $t \sim 1/a^{2}$ to increase by a factor 4 and $U_{H} \sim 1 / R$ to increase by a factor 2. The scaled Hubbard ratio is then $U_{H} / t \sim 40$. By tuning the value of $W$ from zero to $\SI{30}{\milli \electronvolt}$ we can turn bilayer graphene from a non-interacting gas of delocalised electrons to a strongly interacting, highly localised electron system.

\begin{figure}[h]
\caption{Dispersion of the first hole-like bands for different lattice symmetries. These replace the single, flat band in Fig. \ref{fig:CBLG_dispEvo}c. All parameters are the same as in Fig. \ref{fig:CBLG_dispEvo}: $W = \SI{30}{\milli \electronvolt}$, $a = \SI{80}{\nano \meter}$, and $\Delta = \SI{15}{\milli \electronvolt}$. (a) Band $n = -1$ for a triangular lattice potential. (b) Bands $n = -1, \ -2, \ -3$ for a kagome lattice potential. (c) Bands $n = -1, \ -2, \ -3$ for a Lieb lattice potential. Each inset plots the potential energy at the gate in real space, as well as the Brillouin zone. Vertical scale is $\SI{0.8}{\milli \electronvolt}$ in each panel.}
\centering
\includegraphics[width=0.45\textwidth]{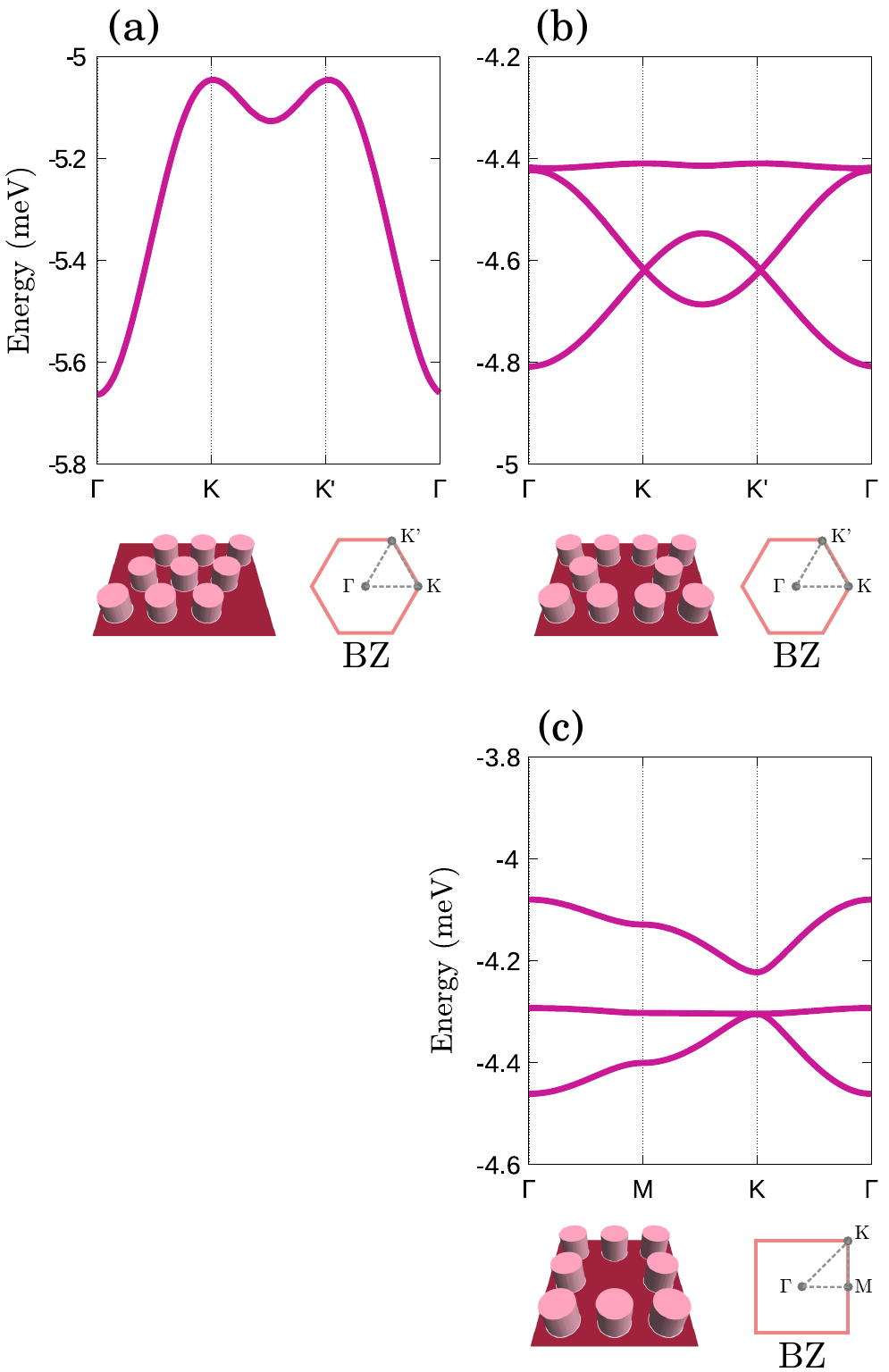}
\label{fig:CLGB_otherSymmetriesDisp}
\end{figure}

\section{Other lattice symmetries}

So far we have considered only the simplest possible patterning, a square array of anti-dots. We have shown that electrons in bilayer graphene can be localised to anti-dot sites by a moderate applied potential. The obvious next step is to try other kinds of anti-dot arrays, and to see whether we can achieve localisation on, say, triangular, kagome, or Lieb lattices. Any of these patterns are experimentally possible.

Figure \ref{fig:CLGB_otherSymmetriesDisp} plots the dispersion of the flat bands for three different lattice symmetries. In each calculation we have kept the parameters identical to those in Fig. \ref{fig:CBLG_dispEvo}c; that is, $W = \SI{30}{\milli \electronvolt}$, $\Delta = \SI{15}{\milli \electronvolt}$ and $a = \SI{80}{\nano \meter}$. Panel (a) plots the first hole band ($n = -1$) for a triangular anti-dot lattice. This band replaces the single flat band in Fig. \ref{fig:CBLG_dispEvoFlatBand} that we computed for the square lattice. We find that the dispersion of the flat band matches that of a tight-binding model on a triangular lattice with nearest-neighbour hopping parameter $t = \SI{0.067}{\milli \electronvolt}$.

We can also reproduce the dispersion of a kagome lattice and a Lieb lattice. The potentials are plotted around one unit cell in the insets of Fig. \ref{fig:CLGB_otherSymmetriesDisp}b and \ref{fig:CLGB_otherSymmetriesDisp}c. In these calculations we have kept the nearest-neighbour distance the same, at $a = \SI{80}{\nano \meter}$, however the overall lattice constant for these patterns is double that. Formally, the Brillouin zones in Fig. \ref{fig:CLGB_otherSymmetriesDisp}b and Fig. \ref{fig:CLGB_otherSymmetriesDisp}c are not the same as those in Fig. \ref{fig:CLGB_otherSymmetriesDisp}a and \ref{fig:CBLG_dispEvoFlatBand}. Since the kagome and the Lieb patterns each have three anti-dot sites per unit cell, the corresponding tight-binding models each have three energy bands. In our calculation for bilayer graphene we thus expect three bands instead of one; the single, isolated flat band in Fig. \ref{fig:CBLG_dispEvo} becomes a triplet of isolated bands in figures \ref{fig:CLGB_otherSymmetriesDisp}b and \ref{fig:CLGB_otherSymmetriesDisp}c whose total bandwidth is roughly the same as the single flat band in Fig. \ref{fig:CBLG_dispEvo}. We find that the dispersion for a kagome anti-dot lattice (Fig. \ref{fig:CLGB_otherSymmetriesDisp}b) mimics that of a kagome tight-binding model with effective hopping parameter $t \approx \SI{0.033}{\milli \electronvolt}$. And the dispersion for a Lieb anti-dot lattice (Fig. \ref{fig:CLGB_otherSymmetriesDisp}c) matches that of a Lieb tight-binding model with the same value $t \approx \SI{0.033}{\milli \electronvolt}$. Note that these dispersions do not map perfectly onto a nearest-neighbour tight-binding model, there is an additional, small, next-nearest-neighbour hopping term, $t'$.

In addition to these patterns one could also create an artificial honeycomb lattice. This would produce two graphene-like bands. Or, to mimic hexagonal boron nitride, one could create a honeycomb lattice with broken sublattice symmetry; for example, by giving each sublattice have a different anti-dot radius. Given the existence of high-quality devices based on real graphene, one may ask why there is a need to engineer artificial graphene. The artificial system gives an opportunity to study the strongly correlated regime, which does not exist in natural graphene.

\section{Comparison with patterned monolayer graphene}

We can demonstrate that bilayer graphene is uniquely suited for this kind of artificial crystal by repeating our calculations for patterned monolayer graphene. In this case the Hamiltonian will be

\begin{align*}
    H =
    \begin{bmatrix}
        0 & v p_{-} \\
        v p_{+} & 0
    \end{bmatrix}
    +
    U(\bm{r})
\end{align*}

Which can be numerically diagonalised by almost exactly the same technique that we described in section \ref{sec:CBLG_theoreticalTechniques}.

\begin{figure}[t]
\caption{Dispersion of patterned monolayer graphene for $W = \SI{0}{\milli \electronvolt}$ and $W = \SI{60}{\milli \electronvolt}$. All other parameters are identical to those used for computing Fig. \ref{fig:CBLG_dispEvo}: $a = \SI{80}{\nano \meter}$, $R = \SI{15}{\nano \meter}$ and $z = \SI{10}{\nano \meter}$}
\centering
\includegraphics[width=0.40\textwidth]{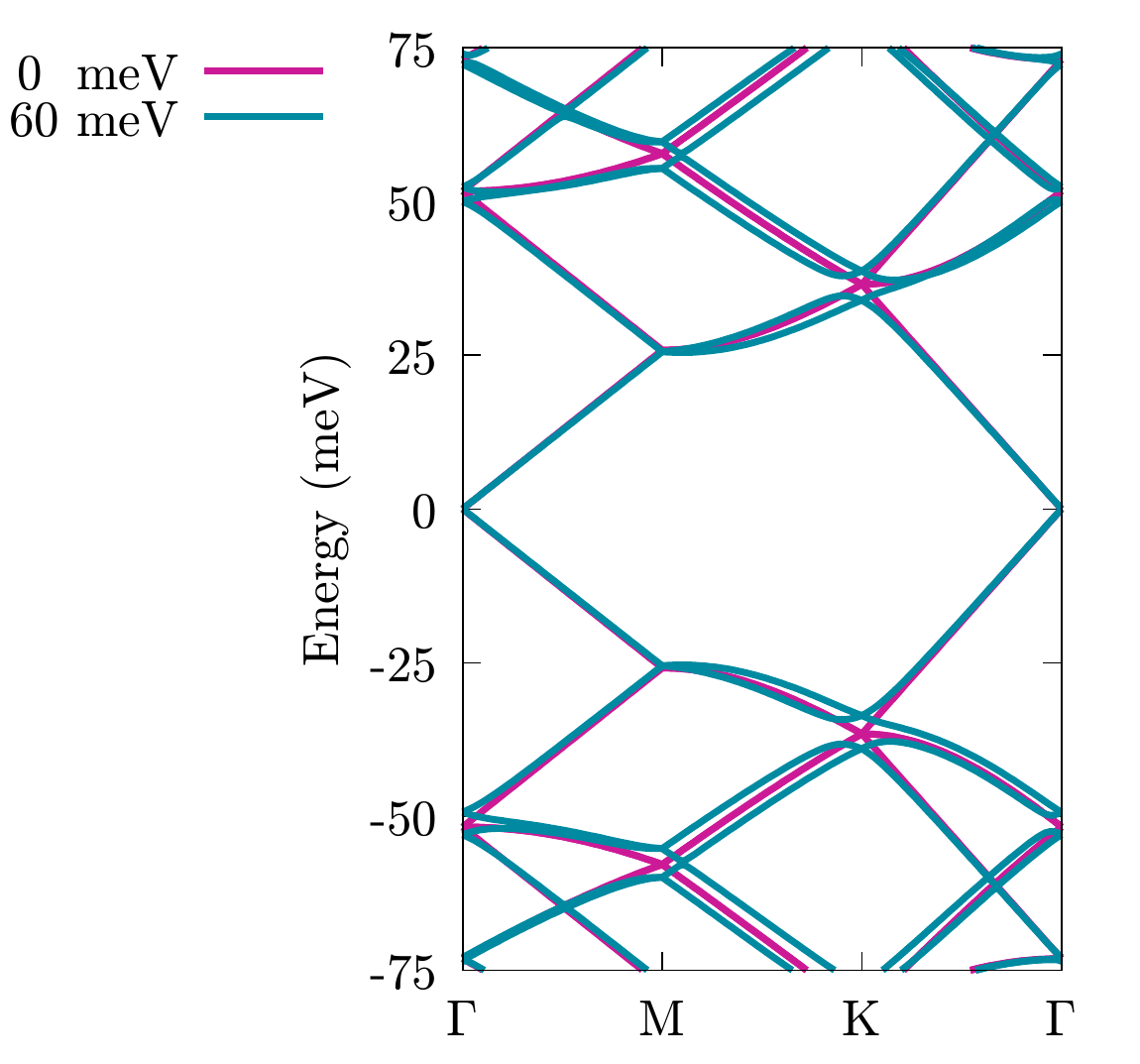}
\label{fig:CBLG_dispEvoMLG}
\end{figure}

The results for patterned monolayer graphene a given in Fig. \ref{fig:CBLG_dispEvoMLG}. To be able to compare this directly with our results for bilayer graphene we have kept all parameters identical to those used for Fig. \ref{fig:CBLG_dispEvo}: the only difference is that two, stacked sheets of graphene have been replaced by one sheet only. Clearly, over the same range of $W$, the linear bands of monolayer graphene are not heavily reconstructed by the periodic potential. Even extending the calculation to $W = \SI{60}{\milli \electronvolt}$ does not provide a significant change in the band structure. The reason for this is partly the difference in energy scales between MLG and BLG, and partly because MLG has no band gap. If $K$ is the magnitude of momentum at the artificial Brillouin-zone vertex then the bare MLG energy at this point is $v K$, and the bare BLG energy at the same point is roughly $K^{2} / 2 m^{*}$. To significantly restructure the energy bands the applied potential has to be larger than each of these energy scales. Since $v K \gg K^{2} / 2 m^{*}$, MLG requires a much larger potential strength. We have also shown that the development of the flat band in BLG is related to the band gap, $\Delta$. The fact that no such equivalent gap exists in MLG is another reason why it is difficult to localise electrons in this system.

\section{New method for addressing Coulomb screening in bilayer graphene}

The amplitude of the imposed potential is reduced by electron-electron screening, and since our results depend on a sufficiently large potential amplitude it is necessary to check that screening doesn't kill the effects described above, entirely. While this issue is practically important, our approach to screening in bilayer graphene has a broader theoretical importance; the technique that we develop is applicable to bilayer graphene in general, and not just to the patterned system we focus on here.

Suppose that we have an imposed potential, $U_{0}(\bm{r})$, and a self-consistent potential, $\widetilde{U}(\bm{r})$, which accounts for screening due to the filled Dirac sea. In general, $\widetilde{U} < U_{0}$, and we must check, for an experimentally realistic $U_{0}$, that the corresponding $\widetilde{U}$ is capable of generating a flat band. We need to address both the shape of $\widetilde{U}(\bm{r})$ and its total amplitude. Our approach is to use the self-consistent Hartree equation

\begin{align}\label{equ:CBLG_hartreeEquation}
    \widetilde{U}(\bm{q})
    =
    U_{0}(\bm{q})
    +
    \frac{2 \pi e^{2}}{\ve q} n_{q}
    \\
    n_{q} = \int
    e^{i \bm{q} \cdot \bm{r}}
    n(\bm{r})
    d^{2} \bm{r}
    \nonumber
\end{align}

Where $n_{q}$ is the Fourier amplitude for the density of particles, a functional of $\widetilde{U}$. Fourier amplitudes of $\widetilde{U}$ and $U_{0}$ are defined similarly. This equation is just an expression of Coulombs law and applies equally well to strongly and weakly correlated systems. The major challenge in applying Eqn. \ref{equ:CBLG_hartreeEquation} equation is to compute the particle density, $n_{q}$, which has contributions from the entire, filled Dirac sea. The band theory that we use here is based on a continuum model of bilayer graphene. Within this model the Dirac sea is formally infinite, the dispersion is unbounded from below.

There are a number of standard approaches to screening which, it turns out, are not applicable to our case. The two major problems are the strong potential modulation, and the infinite Dirac sea; none of the available techniques can address both of these cases simultaneously. In a system with a single parabolic band -- for example, a semiconductor 2DEG -- the total particle density can be computed by brute-force numerics. Here, after numerical diagonalisation of the Hamiltonian, one can explicitly sum over all states from the bottom of the band (at $\ve = 0$) up to the chemical potential ($\ve = \mu$). This technique can address strong external potentials, but does not work if the dispersion extends down to $\ve = -\infty$.

The standard approach to Coulomb screening is the perturbation theory RPA method. The screened potential is given by

\begin{equation*}
    \widetilde{U}_{q} =
    \includegraphics[width=0.32\textwidth,valign=t,raise=-0.10\baselineskip]{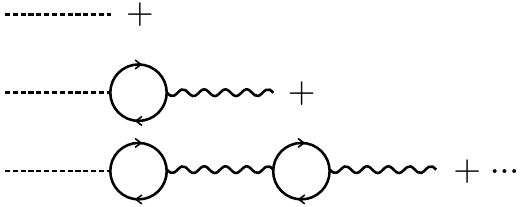}
\end{equation*}

Which is valid in the limit $U^{0} \rightarrow 0$. Here, the dashed represent the external potential, the wavy line represents the Coulomb interaction, and the solid lines represents the electron Green's function. Using this method one can address a system whose dispersion is either bounded from below, such as the semiconductor 2DEG, or unbounded, such as bilayer graphene. One cannot, however, address a system with strong modulation, and this is precisely the limit we wish to consider.

Techniques to account for strong external fields have been developed. The Heisenberg-Euler method is one of these \cite{keser_nonlinear_2022}. That techniques is applicable only to fields which are not spatially varying, and is thus not appropriate to the present case.

In the end, a new technique is required to address Coulomb screening of a strong, spatially varying, external potential in a system with a dispersion unbounded from below. We present the details of our technique in the following section.

\subsection{Description of technique}\label{sec:descriptionOfTechnique}

Since we are working in a continuum model, the Dirac sea is formally infinite. Band theory, as described above, can compute the particle density due to states between $\mu$ and some cut-off energy $-\Lambda$, deep within the Dirac sea; we call this contribution the Bloch contribution. To compute the total particle density it is not sufficient to consider only the Bloch contribution. There is a significant contribution to $n_{q}$ which comes from the bottom of the bilayer graphene band, that is, from states below $-\Lambda$. We call this second contribution to $n_{q}$ the semi-classical contribution, it can be accounted for using a closed-form semi-classical expression.

We account for the semi-classical contribution within the Thomas-Fermi approximation, whose validity is determined by the value of the free parameter $\Lambda$. If the energy $\ve = - \Lambda$ is large enough to treat position and momentum as commuting variables, then the particle density due to states below $- \Lambda$ can be computed semi-classically. Thus, the wavelength which corresponds to $\Lambda$ defined by $\lambda = 2 \pi / \sqrt{2 m^{*} \Lambda}$ should be much smaller than the wavelength, $a$, of the applied potential. We can write this condition as $\lambda \ll a$ or $\Lambda \gg g^{2} / 2 m^{*}$. Note that $x$ and $p$ cannot be treated as commuting variables when dealing with the states above $-\Lambda$; that is, when computing the Bloch contribution.

\begin{figure}[t]
\caption{Schematic illustrating the various contributions to particle density. The Bloch contribution is due to states between $\mu$ and $-\Lambda$, it is captured by our band-theory calculation. The semi-classical contribution is due to states between $-\Lambda$ and the bottom of the band, it is exactly anti-phase with the potential and can be described within the Thomas-Fermi approximation.}
\centering
\includegraphics[width=0.35\textwidth]{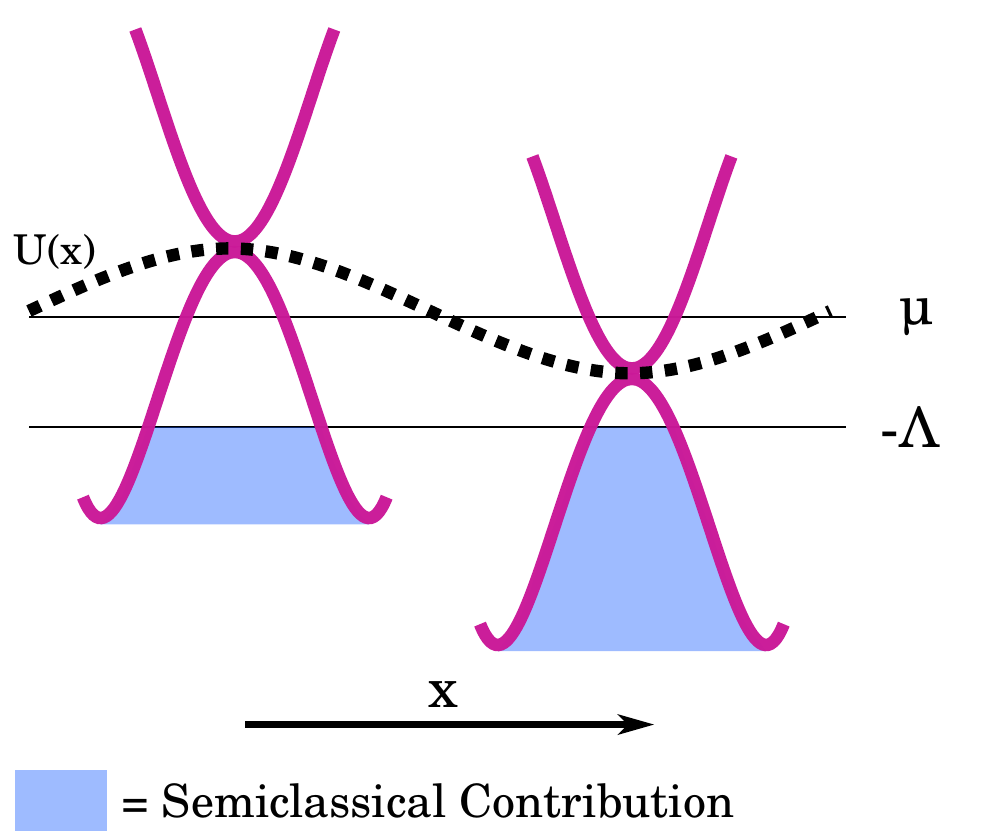}
\label{fig:CBLG_shiftingBand}
\end{figure}

The basic features of the semi-classical contribution can be understood by considering the potential energy in the long wavelength limit as a spatially varying shift in the zero of the bilayer energy bands. This is shown in figure \ref{fig:CBLG_shiftingBand}. The bare bilayer graphene bands are drawn in purple; here, the bottom of the band is indicated schematically by the change in band curvature. The zero of these bands is pinned to the potential energy function, represented schematically by the dashed line. Figure \ref{fig:CBLG_shiftingBand} shows that the states between $-\Lambda$ and the bottom of the band contribute to the spatially varying part of the total particle density. At the minimum of $U(\bm{r})$ the number of states between $-\Lambda$ and the bottom of the band reaches a maximum. Conversely, at the maximum of $U(\bm{r})$ the number of states between $-\Lambda$ and the bottom of the band reaches a minimum. We thus expect that the semi-classical contribution to the particle density has the same periodicity as the potential energy but with the opposite sign.

The above picture can be made quantitative. First, the subsystem we consider (the states drawn in blue in Fig. \ref{fig:CBLG_shiftingBand}) is filled from the very bottom of the band up to the point $\ve = -\Lambda$. Although the true Fermi energy lies close to the charge neutrality point, for the sake of computing the semi-classical contribution to $n(\bm{r})$ we can treat the cut-off, $-\Lambda$, as though it were a kind of Fermi energy. Within the Thomas-Fermi approximation the radius of the constant energy contour, $\ve = -\Lambda$, varies in space. We treat this as a spatially varying ``Fermi momentum", $p(\bm{r})$, for the set of states below $\ve = -\Lambda$. At each point in space the ``Fermi momentum'' is defined by

\begin{align*}
    - \Lambda &= -\frac{p^{2}(\bm{r})}{2m^{*}} + U(\bm{r})
    \\
    \implies
    p^{2}(\bm{r}) &= 2 m^{*} \Lambda + 2 m^{*} U(\bm{r})
\end{align*}

Here, we approximate the bilayer dispersion by a quadratic with effective mass $m^{*}$, valid for sufficiently large $\Lambda$. We want to compute the contribution to particle density due to states below $\ve = - \Lambda$, for which the momentum radius $p(\bm{r})$ defines a region in momentum space absent of any electrons. The total semi-classical contribution will thus be a large, constant part minus a smaller, spatially varying part due to $p(\bm{r})$. That is, $n_{S}(\bm{r}) = const. - 4 \pi p(\bm{r})^{2} / (2 \pi)^{2}$. At each point, $\bm{r}$, in space we count the number of states ``missing'' from the semi-classical part and subtract this from the total number of states in the filled Brillouin zone of the real crystal. Note that if the amplitude of the potential were zero, then the momentum radius would become constant, $p(\bm{r}) = p$, and the Bloch contribution could be computed as the number of states \emph{within} the radius, $p$; that is, $n_{B}(\bm{r}) = n_{B} = 4 \pi p^{2} / (2 \pi)^{2}$. This would exactly cancel with the negative contribution in $n_{S} = const. - 4 \pi p^{2} / (2 \pi)^{2}$ and give the total density as the number of particles in one, fully filled Brillouin zone. When dealing with a non-zero potential amplitude, however, the Bloch contribution cannot be computed using $n_{B}(\bm{r}) = 4 \pi p(\bm{r})^{2} / (2 \pi)^{2}$ (because $\bm{p}$ and $\bm{r}$ do not commute) and hence does not exactly cancel with the semi-classical contribution in this way. In the end, the constant contribution to $n_{S}(\bm{r})$ is inert and we need only consider the spatially varying part, which is

\begin{align*}
    n_{S}(\bm{r}) =
    - \frac{4}{(2 \pi)^{2}} \pi p(\bm{r})^{2} &=
    - \frac{4 \pi}{(2 \pi)^{2}} 2 m^{*} U(\bm{r}) \\ &=
    - \frac{2 m^{*}}{\pi} U(\bm{r})
\end{align*}

This is just the density of states -- accounting for spin and valley degeneracy -- multiplied by the potential energy, with the opposite sign. Since the BLG dispersion contains a higher-energy band starting at around $\ve \approx -\gamma$, we require that $\Lambda$ be small enough to avoid entering this band. For the parameters that we use, the cut-off $\Lambda$ is thus constrained according to $\SI{2}{\milli \electronvolt} \ll \Lambda \ll \SI{380}{\milli \electronvolt}$. In practise, our band theory calculation breaks down when $\Lambda$ is too large, and this occurs well before $\Lambda \sim \gamma$. In what follows we use everywhere $\Lambda = \SI{25}{\milli \electronvolt}$.

The remaining contribution to the particle density is the Bloch contribution, which is computed using

\begin{align*}
    n_{B}(\bm{r}) =
    \sum_{ \substack{ n, \bm{k} \\ -\Lambda < \ve_{n,\bm{k}} < \mu } }
    |\psi_{n,\bm{k}}(\bm{r})|^{2}
\end{align*}

That is, we sum over all Bloch states from $\ve = - \Lambda$ to $\ve = \mu$. The total particle density is the sum of these two contributions

\begin{align}\label{equ:CBLG_densityNumerics}
    n(\bm{r}) =
    n_{B}(\bm{r}) +
    n_{S}(\bm{r})
\end{align}

Each of these two terms needs to be computed using $U(\bm{r}) = \widetilde{U}(\bm{r})$. This technique for computing the particle density (a necessary input for the Hartree equation) is more general than the artificial crystal problem we are studying here. One could just as well apply this method to study, for example, impurity screening in bilayer graphene. The main challenge to overcome is to account for the infinity of states within the Dirac sea, and we do this by combining a band theory calculation with a semi-classical calculation.

\begin{figure}[h]
    \centering
    \caption{Comparison between two calculations of the total variation in particle density for a weak potential. The first calculation, $\delta n_{B} + \delta n_{S}$, is the density variation computed by our method as described in section \ref{sec:descriptionOfTechnique}. The second calculation, $\delta n$, comes from RPA perturbation theory (Eqn. \ref{equ:CBLG_densityPerturbationTheory}).}
    \input{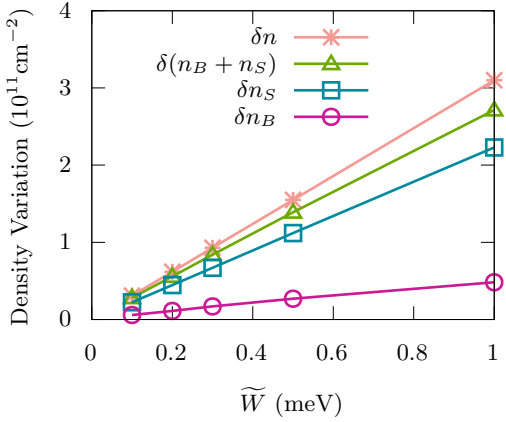}
    \label{fig:CBLG_screeningTest}
\end{figure}

\subsection{Testing density equation against perturbation theory}

We can test the above calculation of $n(\bm{r})$ at $\Delta = 0$ using perturbation theory. We work to linear order in the applied potential, $U^{0}(\bm{r})$, and sum over all orders of the Coulomb RPA chain. Explicitly, perturbation theory, which operates in the plane-wave basis, is valid for $W_{0} \ll g^{2} / 2 m^{*} \sim \SI{2}{\milli \electronvolt}$. This ensures that an energy gap induced by the potential is much smaller than the energy of a plane-wave near the gap.

Suppose that a probe charge with density $\rho(\bm{r})$ is added to a bilayer graphene sheet at charge neutrality. Suppose also that there is an external potential $U^{0}(\bm{r})$. The probe charge interacts both with the external potential and with the particle-density, $n(\bm{r})$, of electrons in the filled Dirac sea. This interaction energy and its corresponding diagram chain are given by

\begin{equation}\label{equ:CBLG_screeningFeynmanDiagrams}
    U^{0}_{q} \rho_{q} + \frac{2 \pi e^{2}}{\ve q} n_{q} \rho_{q} =
\includegraphics[width=0.3\textwidth,valign=t,raise=-0.10\baselineskip]{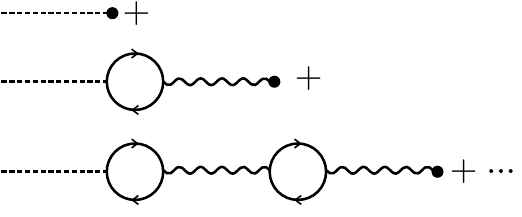}
\end{equation}

Where the dashed line represents the external potential $U^{0}_{q}$, the solid line represents the electron Green's function, the wavy line represents the Coulomb interaction, and the black dot represents the probe charge. Since $U^{0}_{q} \rho_{q}$ corresponds to the first diagram in Eqn. \ref{equ:CBLG_screeningFeynmanDiagrams}, the remaining diagrams can be summed to obtain an expression for $(2 \pi e^{2} / \ve q) n_{q} \rho_{q}$. We find that

\begin{align*}
    \frac{2 \pi e^{2}}{\ve q} n_{q} \rho_{q} =
    U^{0}_{q} \Pi_{q}
    \frac{1}{1 - \frac{2 \pi e^{2}}{\ve q} \Pi_{q}}
    \frac{2 \pi e^{2}}{\ve q} \rho_{q}
\end{align*}

Where $\Pi_{q}$ is the static polarisation operator, corresponding to the electron loop in Eqn. \ref{equ:CBLG_screeningFeynmanDiagrams}. We can solve this equation for $n_{q}$.

\begin{align}\label{equ:CBLG_densityPerturbationTheory}
    n_{q} =
    \frac{U^{0}_{q}}{1 - \frac{2 \pi e^{2}}{\ve q} \Pi_{q}}
    \Pi_{q}
    = \widetilde{U}_{q} \Pi_{q}
\end{align}

In the end we have written this in terms of the self-consistent potential energy $\widetilde{U}$. We are concerned mainly with momenta $|\bm{q}| \sim g$, over which the polarisation operator \cite{hwang_screening_2008} is given by the constant

\begin{align*}
    \Pi_{q} = - 2 \ln(4) \frac{m^{*}}{\pi}
\end{align*}

In this expression for the polarisation operator the entire, filled Dirac sea has been accounted for. We thus have two different ways of computing the particle density: we can use $(n_{B})_{q} + (n_{S})_{q}$, or we can use $- 2 \ln(4) m^{*} \widetilde{U}_{q} / \pi$. Once we have verified that our numerical scheme works in this limit we can apply it to more general situations. For the sake of comparing these two methods we consider a periodic potential with only one Fourier harmonic

\begin{align*}
    \widetilde{U}(\bm{r}) =
    2 \widetilde{W} ( \cos(gx) + \cos(gy) )
\end{align*}

We also neglect the distance between the gate and the 2DEG ($z = 0$). In this case a `weak' potential is defined by $\widetilde{W} \ll \SI{2}{\milli \electronvolt}$ (the equivalent energy scale required for the flat band in Fig. \ref{fig:CBLG_dispEvo}c is $\widetilde{W} \approx \SI{1.3}{\milli \electronvolt}$). Using the above expression for $\widetilde{U}(\bm{r})$ we can compute the resulting particle density using Eqn. \ref{equ:CBLG_densityPerturbationTheory}, or using Eqn. \ref{equ:CBLG_densityNumerics}. We find that the shape of the particle density is always sinusoidal (given a weak, sinusoidal $\widetilde{U}(\bm{r})$), and that the phase of the particle density is the same for each calculation. To compare the two different schemes we can thus consider the quantity $\delta n$ which is the total variation in density from minimum to maximum. The comparison is given in Fig. \ref{fig:CBLG_screeningTest}. Here, we find that $\delta (n_{B} + n_{S})$, computed using Eqn. \ref{equ:CBLG_densityNumerics}, is equal, within $10\%$, to  $\delta n$, computed using Eqn. \ref{equ:CBLG_densityPerturbationTheory}. The equality of these two schemes is non-trivial: the numerical answer was obtained in the Bloch basis, while the analytical answer was obtained in the plane-wave basis. The discrepancy may be accounted for by our use of the parabolic band approximation. This was used in computing $\delta n$ and $\delta n_{S}$, but not in computing $\delta n_{B}$.


We conclude that the scheme for computing particle density, $n(\bm{r}) = n_{B}(\bm{r}) + n_{S}(\bm{r})$, is correct in the limit of weak potential modulation and zero energy gap. In the following section we apply this technique to the more general flat band system.

\begin{figure}[h]
\caption{Plot of $\widetilde{U}(\bm{r})$, defined by Eqn. \ref{equ:CBLG_potentialAtBLG} with $\widetilde{W} = \SI{30}{\milli \electronvolt}$. Also show is the function $U_{0}(\bm{r})$ consistent with the Hartree equation (Eqn. \ref{equ:CBLG_hartreeEquation}). In the first panel the cut has been taken along $y = 0$, and in the second panel the cut is along $y = a/2$. In both panels $\widetilde{U}$ has been scaled by a factor $4.3$.}
\centering
\includegraphics[width=0.35\textwidth]{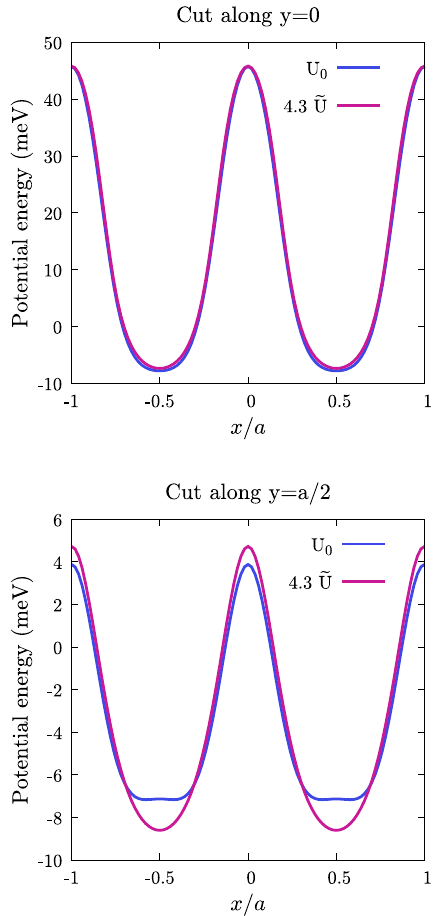}
\label{fig:CBG_screeningU0UTilde}
\end{figure}

\subsection{Hartree screening in the flat band limit}

Having established that the particle density can be computed using Eqn. \ref{equ:CBLG_densityNumerics} we can now compute $n(\bm{r})$ for the flat band system shown in Fig. \ref{fig:CBLG_dispEvo}c. This allows us to find the relationship between $U_{0}$ and $\widetilde{U}$ via the Hartree equation (Eqn. \ref{equ:CBLG_hartreeEquation}). We will work under the assumption that an applied potential, $U_{0}$, will maintain the same shape after screening has been accounted for -- that is, $\widetilde{U}(\bm{r})$ differs from $U_{0}(\bm{r})$ by a constant scaling factor. Our goal is to estimate this scaling factor, which corresponds to a reduction in the effective value of $W$. To show that this assumption is reasonable we solve the Hartree equation in reverse. First, suppose that $\widetilde{U}(\bm{r})$ has exactly the same form as the potential energy defined by Eqn. \ref{equ:CBLG_potentialAtBLG}; its amplitude is $W \mapsto \widetilde{W}$. We take this as an ansatz solution to the Hartree equation. Given the form of $\widetilde{U}(\bm{r})$ we can compute $U_{0}(\bm{r})$ using the Hartree equation

\begin{align*}
    U_{0}(\bm{r}) =
    \widetilde{U}(\bm{r}) -
    \int \frac{e^{2}}{\ve |\bm{r} - \bm{r}'|} n(\bm{r}') d^{2}\bm{r}'
\end{align*}

Where $n(\bm{r})$ is a functional of $\widetilde{U}(\bm{r})$ given by Eqn. \ref{equ:CBLG_densityNumerics}. If this form for $\widetilde{U}(\bm{r})$ is correct then $U_{0}(\bm{r})$ computed using the above equation will have the same form as the original potential; that is, $U_{0}(\bm{r})$ will differ from $\widetilde{U}(\bm{r})$ by a multiplicative factor. We can then characterise the strength of screening in bilayer graphene by estimating this factor.

Indeed, computing $U_{0}(\bm{r})$ in this way gives a function which is essentially of the same form as $\widetilde{U}(\bm{r})$. We present a comparison of these two functions in figure \ref{fig:CBG_screeningU0UTilde} for $\widetilde{W} = \SI{30}{\milli \electronvolt}$ and for $\mu$ at the charge-neutrality point ($f = 0$). In the first panel, which is a cut of both functions along $y = 0$, we find that the two functions are identical up to a scaling factor $4.3$. In the second panel, a cut along $y = a/2$ using the same scaling factor, the two functions are roughly equivalent except for the structure of their minima. We conclude that screening has only a small effect on the shape of the potential, and that its main effect is to scale the amplitude of the potential by a constant factor.

\begin{figure}[t]
\caption{Applied potential strength, $W_{0}$, as a function of the self-consistent potential strength, $\widetilde{W}$, for two different filling fractions, $f$. A filling fraction $f = 0$ corresponds to the charge neutrality point and a filling fraction of $f = - 3n_{0} / 4$, or one quarter filling of the flat band, corresponds to the Mott point.}
\centering
\includegraphics[width=0.4\textwidth]{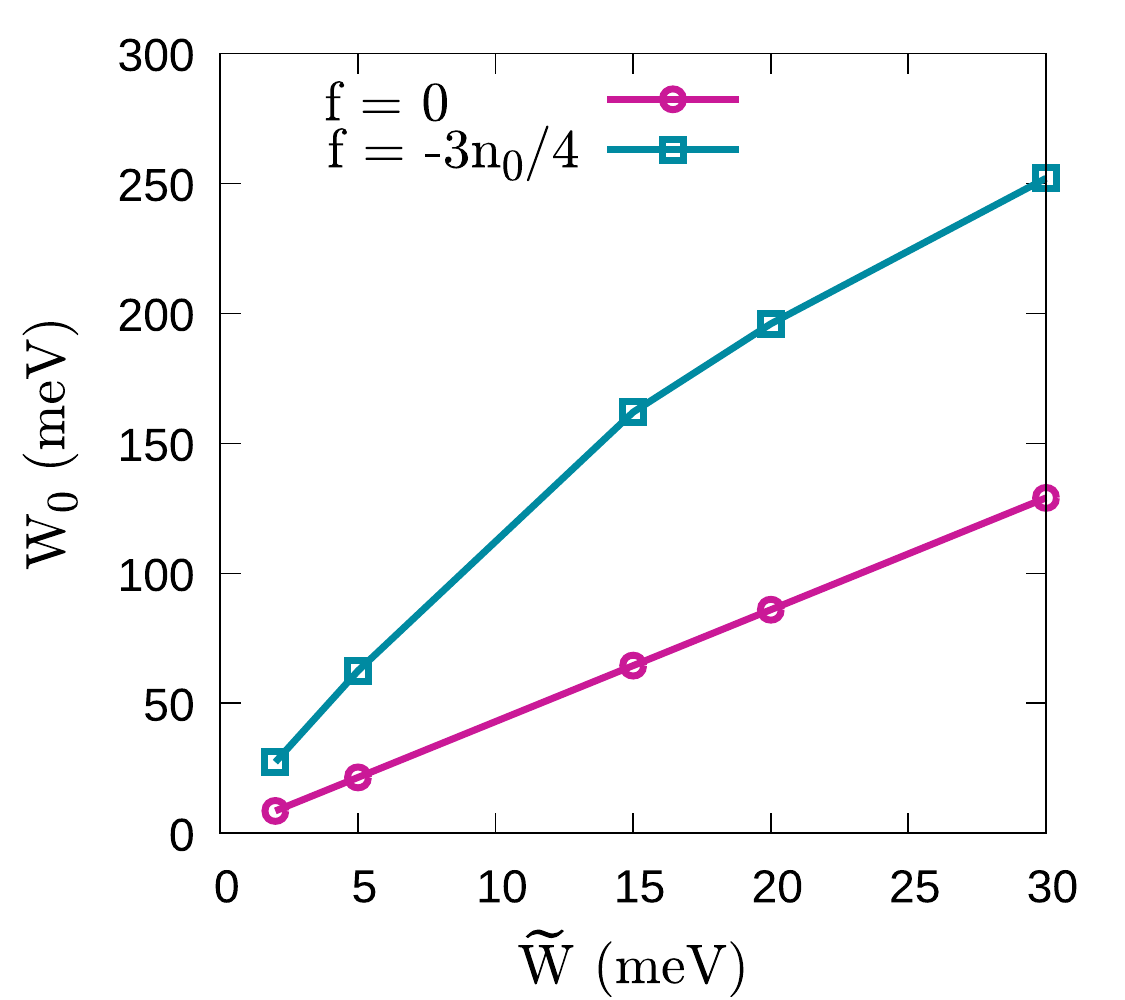}
\label{fig:CBLG_screeningW0}
\end{figure}

We can now estimate the strength of electron-electron screening in patterned bilayer graphene. We focus on two points of interest: the charge-neutrality point and the Mott point (one quarter filling of the flat band, or one electron per anti-dot site). These correspond, respectively, to the point of weakest screeening and to the point of strongest screening (as far as experimentally interesting points within the flat band go). Note that typically the Mott point occurs at half-filling of a band. Here, because the system is both valley and spin degenerate, we must consider quarter-filling. At these two filling fractions we fix the self-consistent potential amplitude, $\widetilde{W}$, and compute the amplitude, $W_{0}$, that the applied potential must have to achieve this value of $\widetilde{W}$. Our results are given in Fig. \ref{fig:CBLG_screeningW0}. At the charge-neutrality point (purple curve) we find that $W_{0}$ is consistently around $4.3$ times larger than $\widetilde{W}$. This means that the effect of screening is to reduce the applied potential by a factor $\approx 4.3$. At the Mott point (blue curve) screening is stronger: to achieve $\widetilde{W} = \SI{30}{\milli \electronvolt}$ an applied potential $W_{0} \approx 8.3 \widetilde{W}$ is required. As the electron density is tuned into the flat band the influence of an applied potential becomes weaker, and the amplitude of this potential needs to be increased to maintain the flat band.

The stronger screening at $f = - 3 n_{0} / 4$ can be understood in terms of the particle density within the flat band that was presented in Fig. \ref{fig:CBLG_chargeDensityCutMap}. The major contribution to the spatial variation of $n_{B}(\bm{r})$ comes from the flat band. This contribution is in phase with the potential, meaning that its sign is opposite to that of $n_{S}(\bm{r})$. It thus acts to reduce the strength of the screening. If we then move $\mu$ to the Mott point we remove a majority of these electrons, which are in-phase with the potential, leading to an increase in the strength of screening.

While the estimate given here is rough, it gives an idea of the strength of screening in these systems. In particular it shows that $\widetilde{W} = \SI{30}{\milli \electronvolt}$ is well within reach of experiments. A PBG device using hBN as a substrate has a breakdown field of $\SI{0.7}{\volt \per \nano \meter}$.\cite{dean_boron_2010} If we take the thickness of hBN to be $z = \SI{10}{\nano \meter}$ then the maximum potential amplitude is $\SI{7}{\electronvolt}$, well above the value $W_{0} = \SI{250}{\milli \electronvolt}$ required for a fully developed flat band at the point of strongest screening.

\section{Conclusions}

We have suggested a simple method for generating a flat band in bilayer graphene. Within this flat band we find that: (i) Bilayer graphene can be driven into the strongly correlated regime by application of a band gap and long-wavelength, periodic electric field. (ii) The strongly correlated regime can be accessed by either tuning the band gap or the amplitude of the periodic modulation, keeping electron density fixed. (iii) We present a general technique for addressing Hartree screening in bilayer graphene. While this technique has more broad theoretical applications, we use it to verify that electron-electron screening of the periodic potential is not strong enough to destroy the effects listed above. (iv) Patterned bilayer graphene can mimic the dispersions of a variety of different two-dimensional lattices, including square, triangular, kagome, and Lieb lattices. By appropriate patterning it is possible to design the underlying effective Hubbard model.

\begin{acknowledgements}

We wish to acknowledge O. Klochan, who catalysed this project. H. Scammell, F. Xiang, and A. Keser also provided useful comments. This research was supported by an Australian Government Research Training Program (RTP) Scholarship. We have also received support from the Australian Research Council Centre of Excellence in Future Low-Energy Electronics Technology (FLEET) (CE170100039).

\end{acknowledgements}

\bibliography{bib.bib}

\begin{thebibliography}{34}
\expandafter\ifx\csname natexlab\endcsname\relax\def\natexlab#1{#1}\fi
\expandafter\ifx\csname bibnamefont\endcsname\relax
  \def\bibnamefont#1{#1}\fi
\expandafter\ifx\csname bibfnamefont\endcsname\relax
  \def\bibfnamefont#1{#1}\fi
\expandafter\ifx\csname citenamefont\endcsname\relax
  \def\citenamefont#1{#1}\fi
\expandafter\ifx\csname url\endcsname\relax
  \def\url#1{\texttt{#1}}\fi
\expandafter\ifx\csname urlprefix\endcsname\relax\def\urlprefix{URL }\fi
\providecommand{\bibinfo}[2]{#2}
\providecommand{\eprint}[2][]{\url{#2}}

\bibitem[{\citenamefont{Cao et~al.}(2018{\natexlab{a}})\citenamefont{Cao,
  Fatemi, Fang, Watanabe, Taniguchi, Kaxiras, and
  Jarillo-Herrero}}]{cao_unconventional_2018}
\bibinfo{author}{\bibfnamefont{Y.}~\bibnamefont{Cao}},
  \bibinfo{author}{\bibfnamefont{V.}~\bibnamefont{Fatemi}},
  \bibinfo{author}{\bibfnamefont{S.}~\bibnamefont{Fang}},
  \bibinfo{author}{\bibfnamefont{K.}~\bibnamefont{Watanabe}},
  \bibinfo{author}{\bibfnamefont{T.}~\bibnamefont{Taniguchi}},
  \bibinfo{author}{\bibfnamefont{E.}~\bibnamefont{Kaxiras}}, \bibnamefont{and}
  \bibinfo{author}{\bibfnamefont{P.}~\bibnamefont{Jarillo-Herrero}},
  \bibinfo{journal}{Nature} \textbf{\bibinfo{volume}{556}}, \bibinfo{pages}{43}
  (\bibinfo{year}{2018}{\natexlab{a}}), ISSN \bibinfo{issn}{0028-0836,
  1476-4687}, \urlprefix\url{http://www.nature.com/articles/nature26160}.

\bibitem[{\citenamefont{Bistritzer and
  MacDonald}(2011)}]{bistritzer_moire_2011}
\bibinfo{author}{\bibfnamefont{R.}~\bibnamefont{Bistritzer}} \bibnamefont{and}
  \bibinfo{author}{\bibfnamefont{A.~H.} \bibnamefont{MacDonald}},
  \bibinfo{journal}{Proceedings of the National Academy of Sciences}
  \textbf{\bibinfo{volume}{108}}, \bibinfo{pages}{12233}
  (\bibinfo{year}{2011}), ISSN \bibinfo{issn}{0027-8424, 1091-6490},
  \urlprefix\url{https://www.pnas.org/content/108/30/12233}.

\bibitem[{\citenamefont{Lu et~al.}(2019)\citenamefont{Lu, Stepanov, Yang, Xie,
  Aamir, Das, Urgell, Watanabe, Taniguchi, Zhang
  et~al.}}]{lu_superconductors_2019}
\bibinfo{author}{\bibfnamefont{X.}~\bibnamefont{Lu}},
  \bibinfo{author}{\bibfnamefont{P.}~\bibnamefont{Stepanov}},
  \bibinfo{author}{\bibfnamefont{W.}~\bibnamefont{Yang}},
  \bibinfo{author}{\bibfnamefont{M.}~\bibnamefont{Xie}},
  \bibinfo{author}{\bibfnamefont{M.~A.} \bibnamefont{Aamir}},
  \bibinfo{author}{\bibfnamefont{I.}~\bibnamefont{Das}},
  \bibinfo{author}{\bibfnamefont{C.}~\bibnamefont{Urgell}},
  \bibinfo{author}{\bibfnamefont{K.}~\bibnamefont{Watanabe}},
  \bibinfo{author}{\bibfnamefont{T.}~\bibnamefont{Taniguchi}},
  \bibinfo{author}{\bibfnamefont{G.}~\bibnamefont{Zhang}},
  \bibnamefont{et~al.}, \bibinfo{journal}{Nature}
  \textbf{\bibinfo{volume}{574}}, \bibinfo{pages}{653} (\bibinfo{year}{2019}),
  ISSN \bibinfo{issn}{1476-4687},
  \urlprefix\url{https://www.nature.com/articles/s41586-019-1695-0}.

\bibitem[{\citenamefont{Yankowitz et~al.}(2019)\citenamefont{Yankowitz, Chen,
  Polshyn, Zhang, Watanabe, Taniguchi, Graf, Young, and
  Dean}}]{yankowitz_tuning_2019}
\bibinfo{author}{\bibfnamefont{M.}~\bibnamefont{Yankowitz}},
  \bibinfo{author}{\bibfnamefont{S.}~\bibnamefont{Chen}},
  \bibinfo{author}{\bibfnamefont{H.}~\bibnamefont{Polshyn}},
  \bibinfo{author}{\bibfnamefont{Y.}~\bibnamefont{Zhang}},
  \bibinfo{author}{\bibfnamefont{K.}~\bibnamefont{Watanabe}},
  \bibinfo{author}{\bibfnamefont{T.}~\bibnamefont{Taniguchi}},
  \bibinfo{author}{\bibfnamefont{D.}~\bibnamefont{Graf}},
  \bibinfo{author}{\bibfnamefont{A.~F.} \bibnamefont{Young}}, \bibnamefont{and}
  \bibinfo{author}{\bibfnamefont{C.~R.} \bibnamefont{Dean}},
  \bibinfo{journal}{Science}  (\bibinfo{year}{2019}),
  \urlprefix\url{https://www.science.org/doi/abs/10.1126/science.aav1910}.

\bibitem[{\citenamefont{Chen et~al.}(2019{\natexlab{a}})\citenamefont{Chen,
  Sharpe, Gallagher, Rosen, Fox, Jiang, Lyu, Li, Watanabe, Taniguchi
  et~al.}}]{chen_signatures_2019}
\bibinfo{author}{\bibfnamefont{G.}~\bibnamefont{Chen}},
  \bibinfo{author}{\bibfnamefont{A.~L.} \bibnamefont{Sharpe}},
  \bibinfo{author}{\bibfnamefont{P.}~\bibnamefont{Gallagher}},
  \bibinfo{author}{\bibfnamefont{I.~T.} \bibnamefont{Rosen}},
  \bibinfo{author}{\bibfnamefont{E.~J.} \bibnamefont{Fox}},
  \bibinfo{author}{\bibfnamefont{L.}~\bibnamefont{Jiang}},
  \bibinfo{author}{\bibfnamefont{B.}~\bibnamefont{Lyu}},
  \bibinfo{author}{\bibfnamefont{H.}~\bibnamefont{Li}},
  \bibinfo{author}{\bibfnamefont{K.}~\bibnamefont{Watanabe}},
  \bibinfo{author}{\bibfnamefont{T.}~\bibnamefont{Taniguchi}},
  \bibnamefont{et~al.}, \bibinfo{journal}{Nature}
  \textbf{\bibinfo{volume}{572}}, \bibinfo{pages}{215}
  (\bibinfo{year}{2019}{\natexlab{a}}), ISSN \bibinfo{issn}{1476-4687},
  \urlprefix\url{https://www.nature.com/articles/s41586-019-1393-y}.

\bibitem[{\citenamefont{Park et~al.}(2021)\citenamefont{Park, Cao, Watanabe,
  Taniguchi, and Jarillo-Herrero}}]{park_tunable_2021}
\bibinfo{author}{\bibfnamefont{J.~M.} \bibnamefont{Park}},
  \bibinfo{author}{\bibfnamefont{Y.}~\bibnamefont{Cao}},
  \bibinfo{author}{\bibfnamefont{K.}~\bibnamefont{Watanabe}},
  \bibinfo{author}{\bibfnamefont{T.}~\bibnamefont{Taniguchi}},
  \bibnamefont{and}
  \bibinfo{author}{\bibfnamefont{P.}~\bibnamefont{Jarillo-Herrero}},
  \bibinfo{journal}{Nature} \textbf{\bibinfo{volume}{590}},
  \bibinfo{pages}{249} (\bibinfo{year}{2021}), ISSN \bibinfo{issn}{1476-4687},
  \urlprefix\url{https://www.nature.com/articles/s41586-021-03192-0}.

\bibitem[{\citenamefont{Cao et~al.}(2018{\natexlab{b}})\citenamefont{Cao,
  Fatemi, Demir, Fang, Tomarken, Luo, Sanchez-Yamagishi, Watanabe, Taniguchi,
  Kaxiras et~al.}}]{cao_correlated_2018}
\bibinfo{author}{\bibfnamefont{Y.}~\bibnamefont{Cao}},
  \bibinfo{author}{\bibfnamefont{V.}~\bibnamefont{Fatemi}},
  \bibinfo{author}{\bibfnamefont{A.}~\bibnamefont{Demir}},
  \bibinfo{author}{\bibfnamefont{S.}~\bibnamefont{Fang}},
  \bibinfo{author}{\bibfnamefont{S.~L.} \bibnamefont{Tomarken}},
  \bibinfo{author}{\bibfnamefont{J.~Y.} \bibnamefont{Luo}},
  \bibinfo{author}{\bibfnamefont{J.~D.} \bibnamefont{Sanchez-Yamagishi}},
  \bibinfo{author}{\bibfnamefont{K.}~\bibnamefont{Watanabe}},
  \bibinfo{author}{\bibfnamefont{T.}~\bibnamefont{Taniguchi}},
  \bibinfo{author}{\bibfnamefont{E.}~\bibnamefont{Kaxiras}},
  \bibnamefont{et~al.}, \bibinfo{journal}{Nature}
  \textbf{\bibinfo{volume}{556}}, \bibinfo{pages}{80}
  (\bibinfo{year}{2018}{\natexlab{b}}), ISSN \bibinfo{issn}{1476-4687},
  \urlprefix\url{https://www.nature.com/articles/nature26154/}.

\bibitem[{\citenamefont{Kerelsky et~al.}(2019)\citenamefont{Kerelsky, McGilly,
  Kennes, Xian, Yankowitz, Chen, Watanabe, Taniguchi, Hone, Dean
  et~al.}}]{kerelsky_maximized_2019}
\bibinfo{author}{\bibfnamefont{A.}~\bibnamefont{Kerelsky}},
  \bibinfo{author}{\bibfnamefont{L.~J.} \bibnamefont{McGilly}},
  \bibinfo{author}{\bibfnamefont{D.~M.} \bibnamefont{Kennes}},
  \bibinfo{author}{\bibfnamefont{L.}~\bibnamefont{Xian}},
  \bibinfo{author}{\bibfnamefont{M.}~\bibnamefont{Yankowitz}},
  \bibinfo{author}{\bibfnamefont{S.}~\bibnamefont{Chen}},
  \bibinfo{author}{\bibfnamefont{K.}~\bibnamefont{Watanabe}},
  \bibinfo{author}{\bibfnamefont{T.}~\bibnamefont{Taniguchi}},
  \bibinfo{author}{\bibfnamefont{J.}~\bibnamefont{Hone}},
  \bibinfo{author}{\bibfnamefont{C.}~\bibnamefont{Dean}}, \bibnamefont{et~al.},
  \bibinfo{journal}{Nature} \textbf{\bibinfo{volume}{572}}, \bibinfo{pages}{95}
  (\bibinfo{year}{2019}), ISSN \bibinfo{issn}{1476-4687},
  \urlprefix\url{https://www.nature.com/articles/s41586-019-1431-9}.

\bibitem[{\citenamefont{Chen et~al.}(2019{\natexlab{b}})\citenamefont{Chen,
  Jiang, Wu, Lyu, Li, Chittari, Watanabe, Taniguchi, Shi, Jung
  et~al.}}]{chen_evidence_2019}
\bibinfo{author}{\bibfnamefont{G.}~\bibnamefont{Chen}},
  \bibinfo{author}{\bibfnamefont{L.}~\bibnamefont{Jiang}},
  \bibinfo{author}{\bibfnamefont{S.}~\bibnamefont{Wu}},
  \bibinfo{author}{\bibfnamefont{B.}~\bibnamefont{Lyu}},
  \bibinfo{author}{\bibfnamefont{H.}~\bibnamefont{Li}},
  \bibinfo{author}{\bibfnamefont{B.~L.} \bibnamefont{Chittari}},
  \bibinfo{author}{\bibfnamefont{K.}~\bibnamefont{Watanabe}},
  \bibinfo{author}{\bibfnamefont{T.}~\bibnamefont{Taniguchi}},
  \bibinfo{author}{\bibfnamefont{Z.}~\bibnamefont{Shi}},
  \bibinfo{author}{\bibfnamefont{J.}~\bibnamefont{Jung}}, \bibnamefont{et~al.},
  \bibinfo{journal}{Nature Physics} \textbf{\bibinfo{volume}{15}},
  \bibinfo{pages}{237} (\bibinfo{year}{2019}{\natexlab{b}}), ISSN
  \bibinfo{issn}{1745-2481},
  \urlprefix\url{https://www.nature.com/articles/s41567-018-0387-2}.

\bibitem[{\citenamefont{Sharpe et~al.}(2019)\citenamefont{Sharpe, Fox, Barnard,
  Finney, Watanabe, Taniguchi, Kastner, and
  Goldhaber-Gordon}}]{sharpe_emergent_2019}
\bibinfo{author}{\bibfnamefont{A.~L.} \bibnamefont{Sharpe}},
  \bibinfo{author}{\bibfnamefont{E.~J.} \bibnamefont{Fox}},
  \bibinfo{author}{\bibfnamefont{A.~W.} \bibnamefont{Barnard}},
  \bibinfo{author}{\bibfnamefont{J.}~\bibnamefont{Finney}},
  \bibinfo{author}{\bibfnamefont{K.}~\bibnamefont{Watanabe}},
  \bibinfo{author}{\bibfnamefont{T.}~\bibnamefont{Taniguchi}},
  \bibinfo{author}{\bibfnamefont{M.~A.} \bibnamefont{Kastner}},
  \bibnamefont{and}
  \bibinfo{author}{\bibfnamefont{D.}~\bibnamefont{Goldhaber-Gordon}},
  \bibinfo{journal}{Science}  (\bibinfo{year}{2019}),
  \urlprefix\url{https://www.science.org/doi/abs/10.1126/science.aaw3780}.

\bibitem[{\citenamefont{Serlin et~al.}(2020)\citenamefont{Serlin, Tschirhart,
  Polshyn, Zhang, Zhu, Watanabe, Taniguchi, Balents, and
  Young}}]{serlin_intrinsic_2020}
\bibinfo{author}{\bibfnamefont{M.}~\bibnamefont{Serlin}},
  \bibinfo{author}{\bibfnamefont{C.~L.} \bibnamefont{Tschirhart}},
  \bibinfo{author}{\bibfnamefont{H.}~\bibnamefont{Polshyn}},
  \bibinfo{author}{\bibfnamefont{Y.}~\bibnamefont{Zhang}},
  \bibinfo{author}{\bibfnamefont{J.}~\bibnamefont{Zhu}},
  \bibinfo{author}{\bibfnamefont{K.}~\bibnamefont{Watanabe}},
  \bibinfo{author}{\bibfnamefont{T.}~\bibnamefont{Taniguchi}},
  \bibinfo{author}{\bibfnamefont{L.}~\bibnamefont{Balents}}, \bibnamefont{and}
  \bibinfo{author}{\bibfnamefont{A.~F.} \bibnamefont{Young}},
  \bibinfo{journal}{Science}  (\bibinfo{year}{2020}),
  \urlprefix\url{https://www.science.org/doi/abs/10.1126/science.aay5533}.

\bibitem[{\citenamefont{Jiang et~al.}(2019)\citenamefont{Jiang, Lai, Watanabe,
  Taniguchi, Haule, Mao, and Andrei}}]{jiang_charge_2019}
\bibinfo{author}{\bibfnamefont{Y.}~\bibnamefont{Jiang}},
  \bibinfo{author}{\bibfnamefont{X.}~\bibnamefont{Lai}},
  \bibinfo{author}{\bibfnamefont{K.}~\bibnamefont{Watanabe}},
  \bibinfo{author}{\bibfnamefont{T.}~\bibnamefont{Taniguchi}},
  \bibinfo{author}{\bibfnamefont{K.}~\bibnamefont{Haule}},
  \bibinfo{author}{\bibfnamefont{J.}~\bibnamefont{Mao}}, \bibnamefont{and}
  \bibinfo{author}{\bibfnamefont{E.~Y.} \bibnamefont{Andrei}},
  \bibinfo{journal}{Nature} \textbf{\bibinfo{volume}{573}}, \bibinfo{pages}{91}
  (\bibinfo{year}{2019}), ISSN \bibinfo{issn}{1476-4687},
  \urlprefix\url{https://www.nature.com/articles/s41586-019-1460-4}.

\bibitem[{\citenamefont{Rubio-Verdú et~al.}(2021)\citenamefont{Rubio-Verdú,
  Turkel, Song, Klebl, Samajdar, Scheurer, Venderbos, Watanabe, Taniguchi,
  Ochoa et~al.}}]{rubio-verdu_moire_2021}
\bibinfo{author}{\bibfnamefont{C.}~\bibnamefont{Rubio-Verdú}},
  \bibinfo{author}{\bibfnamefont{S.}~\bibnamefont{Turkel}},
  \bibinfo{author}{\bibfnamefont{Y.}~\bibnamefont{Song}},
  \bibinfo{author}{\bibfnamefont{L.}~\bibnamefont{Klebl}},
  \bibinfo{author}{\bibfnamefont{R.}~\bibnamefont{Samajdar}},
  \bibinfo{author}{\bibfnamefont{M.~S.} \bibnamefont{Scheurer}},
  \bibinfo{author}{\bibfnamefont{J.~W.~F.} \bibnamefont{Venderbos}},
  \bibinfo{author}{\bibfnamefont{K.}~\bibnamefont{Watanabe}},
  \bibinfo{author}{\bibfnamefont{T.}~\bibnamefont{Taniguchi}},
  \bibinfo{author}{\bibfnamefont{H.}~\bibnamefont{Ochoa}},
  \bibnamefont{et~al.}, \bibinfo{journal}{Nature Physics} pp.
  \bibinfo{pages}{1--7} (\bibinfo{year}{2021}), ISSN \bibinfo{issn}{1745-2481},
  \urlprefix\url{https://www.nature.com/articles/s41567-021-01438-2}.

\bibitem[{\citenamefont{Zhang et~al.}(2020)\citenamefont{Zhang, Wang, Watanabe,
  Taniguchi, Ueno, Tutuc, and LeRoy}}]{zhang_flat_2020}
\bibinfo{author}{\bibfnamefont{Z.}~\bibnamefont{Zhang}},
  \bibinfo{author}{\bibfnamefont{Y.}~\bibnamefont{Wang}},
  \bibinfo{author}{\bibfnamefont{K.}~\bibnamefont{Watanabe}},
  \bibinfo{author}{\bibfnamefont{T.}~\bibnamefont{Taniguchi}},
  \bibinfo{author}{\bibfnamefont{K.}~\bibnamefont{Ueno}},
  \bibinfo{author}{\bibfnamefont{E.}~\bibnamefont{Tutuc}}, \bibnamefont{and}
  \bibinfo{author}{\bibfnamefont{B.~J.} \bibnamefont{LeRoy}},
  \bibinfo{journal}{Nature Physics} \textbf{\bibinfo{volume}{16}},
  \bibinfo{pages}{1093} (\bibinfo{year}{2020}), ISSN \bibinfo{issn}{1745-2473,
  1745-2481}, \urlprefix\url{http://www.nature.com/articles/s41567-020-0958-x}.

\bibitem[{\citenamefont{Lin et~al.}(2018)\citenamefont{Lin, Choi, Zhang, Qin,
  Yi, Wang, Li, Wang, Zhang, Sun et~al.}}]{lin_flatbands_2018}
\bibinfo{author}{\bibfnamefont{Z.}~\bibnamefont{Lin}},
  \bibinfo{author}{\bibfnamefont{J.-H.} \bibnamefont{Choi}},
  \bibinfo{author}{\bibfnamefont{Q.}~\bibnamefont{Zhang}},
  \bibinfo{author}{\bibfnamefont{W.}~\bibnamefont{Qin}},
  \bibinfo{author}{\bibfnamefont{S.}~\bibnamefont{Yi}},
  \bibinfo{author}{\bibfnamefont{P.}~\bibnamefont{Wang}},
  \bibinfo{author}{\bibfnamefont{L.}~\bibnamefont{Li}},
  \bibinfo{author}{\bibfnamefont{Y.}~\bibnamefont{Wang}},
  \bibinfo{author}{\bibfnamefont{H.}~\bibnamefont{Zhang}},
  \bibinfo{author}{\bibfnamefont{Z.}~\bibnamefont{Sun}}, \bibnamefont{et~al.},
  \bibinfo{journal}{Physical Review Letters} \textbf{\bibinfo{volume}{121}},
  \bibinfo{pages}{096401} (\bibinfo{year}{2018}),
  \urlprefix\url{https://link.aps.org/doi/10.1103/PhysRevLett.121.096401}.

\bibitem[{\citenamefont{Ortiz et~al.}(2020)\citenamefont{Ortiz, Teicher, Hu,
  Zuo, Sarte, Schueller, Abeykoon, Krogstad, Rosenkranz, Osborn
  et~al.}}]{ortiz_cs_2020}
\bibinfo{author}{\bibfnamefont{B.~R.} \bibnamefont{Ortiz}},
  \bibinfo{author}{\bibfnamefont{S.~M.} \bibnamefont{Teicher}},
  \bibinfo{author}{\bibfnamefont{Y.}~\bibnamefont{Hu}},
  \bibinfo{author}{\bibfnamefont{J.~L.} \bibnamefont{Zuo}},
  \bibinfo{author}{\bibfnamefont{P.~M.} \bibnamefont{Sarte}},
  \bibinfo{author}{\bibfnamefont{E.~C.} \bibnamefont{Schueller}},
  \bibinfo{author}{\bibfnamefont{A.~M.} \bibnamefont{Abeykoon}},
  \bibinfo{author}{\bibfnamefont{M.~J.} \bibnamefont{Krogstad}},
  \bibinfo{author}{\bibfnamefont{S.}~\bibnamefont{Rosenkranz}},
  \bibinfo{author}{\bibfnamefont{R.}~\bibnamefont{Osborn}},
  \bibnamefont{et~al.}, \bibinfo{journal}{Physical Review Letters}
  \textbf{\bibinfo{volume}{125}}, \bibinfo{pages}{247002}
  (\bibinfo{year}{2020}), ISSN \bibinfo{issn}{0031-9007, 1079-7114},
  \urlprefix\url{https://link.aps.org/doi/10.1103/PhysRevLett.125.247002}.

\bibitem[{\citenamefont{Ortiz et~al.}(2021)\citenamefont{Ortiz, Sarte, Kenney,
  Graf, Teicher, Seshadri, and Wilson}}]{ortiz_superconductivity_2021}
\bibinfo{author}{\bibfnamefont{B.~R.} \bibnamefont{Ortiz}},
  \bibinfo{author}{\bibfnamefont{P.~M.} \bibnamefont{Sarte}},
  \bibinfo{author}{\bibfnamefont{E.~M.} \bibnamefont{Kenney}},
  \bibinfo{author}{\bibfnamefont{M.~J.} \bibnamefont{Graf}},
  \bibinfo{author}{\bibfnamefont{S.~M.~L.} \bibnamefont{Teicher}},
  \bibinfo{author}{\bibfnamefont{R.}~\bibnamefont{Seshadri}}, \bibnamefont{and}
  \bibinfo{author}{\bibfnamefont{S.~D.} \bibnamefont{Wilson}},
  \bibinfo{journal}{Physical Review Materials} \textbf{\bibinfo{volume}{5}},
  \bibinfo{pages}{034801} (\bibinfo{year}{2021}), ISSN
  \bibinfo{issn}{2475-9953},
  \urlprefix\url{https://link.aps.org/doi/10.1103/PhysRevMaterials.5.034801}.

\bibitem[{\citenamefont{Chen et~al.}(2021)\citenamefont{Chen, Wang, Yin, Gu,
  Jiang, Tu, Gong, Uwatoko, Sun, Lei et~al.}}]{chen_double_2021}
\bibinfo{author}{\bibfnamefont{K.}~\bibnamefont{Chen}},
  \bibinfo{author}{\bibfnamefont{N.}~\bibnamefont{Wang}},
  \bibinfo{author}{\bibfnamefont{Q.}~\bibnamefont{Yin}},
  \bibinfo{author}{\bibfnamefont{Y.}~\bibnamefont{Gu}},
  \bibinfo{author}{\bibfnamefont{K.}~\bibnamefont{Jiang}},
  \bibinfo{author}{\bibfnamefont{Z.}~\bibnamefont{Tu}},
  \bibinfo{author}{\bibfnamefont{C.}~\bibnamefont{Gong}},
  \bibinfo{author}{\bibfnamefont{Y.}~\bibnamefont{Uwatoko}},
  \bibinfo{author}{\bibfnamefont{J.}~\bibnamefont{Sun}},
  \bibinfo{author}{\bibfnamefont{H.}~\bibnamefont{Lei}}, \bibnamefont{et~al.},
  \bibinfo{journal}{Physical Review Letters} \textbf{\bibinfo{volume}{126}},
  \bibinfo{pages}{247001} (\bibinfo{year}{2021}), ISSN
  \bibinfo{issn}{0031-9007, 1079-7114},
  \urlprefix\url{https://link.aps.org/doi/10.1103/PhysRevLett.126.247001}.

\bibitem[{\citenamefont{Zhu et~al.}(2021)\citenamefont{Zhu, Yang, Xia, Yin,
  Wang, Zhao, Dai, Tu, Song, Tao et~al.}}]{zhu_double-dome_2021}
\bibinfo{author}{\bibfnamefont{C.~C.} \bibnamefont{Zhu}},
  \bibinfo{author}{\bibfnamefont{X.~F.} \bibnamefont{Yang}},
  \bibinfo{author}{\bibfnamefont{W.}~\bibnamefont{Xia}},
  \bibinfo{author}{\bibfnamefont{Q.~W.} \bibnamefont{Yin}},
  \bibinfo{author}{\bibfnamefont{L.~S.} \bibnamefont{Wang}},
  \bibinfo{author}{\bibfnamefont{C.~C.} \bibnamefont{Zhao}},
  \bibinfo{author}{\bibfnamefont{D.~Z.} \bibnamefont{Dai}},
  \bibinfo{author}{\bibfnamefont{C.~P.} \bibnamefont{Tu}},
  \bibinfo{author}{\bibfnamefont{B.~Q.} \bibnamefont{Song}},
  \bibinfo{author}{\bibfnamefont{Z.~C.} \bibnamefont{Tao}},
  \bibnamefont{et~al.}, \bibinfo{journal}{arXiv:2104.14487 [cond-mat]}
  (\bibinfo{year}{2021}), \bibinfo{note}{arXiv: 2104.14487},
  \urlprefix\url{http://arxiv.org/abs/2104.14487}.

\bibitem[{\citenamefont{Jiang et~al.}(2021)\citenamefont{Jiang, Yin, Denner,
  Shumiya, Ortiz, Xu, Guguchia, He, Hossain, Liu
  et~al.}}]{jiang_unconventional_2021}
\bibinfo{author}{\bibfnamefont{Y.-X.} \bibnamefont{Jiang}},
  \bibinfo{author}{\bibfnamefont{J.-X.} \bibnamefont{Yin}},
  \bibinfo{author}{\bibfnamefont{M.~M.} \bibnamefont{Denner}},
  \bibinfo{author}{\bibfnamefont{N.}~\bibnamefont{Shumiya}},
  \bibinfo{author}{\bibfnamefont{B.~R.} \bibnamefont{Ortiz}},
  \bibinfo{author}{\bibfnamefont{G.}~\bibnamefont{Xu}},
  \bibinfo{author}{\bibfnamefont{Z.}~\bibnamefont{Guguchia}},
  \bibinfo{author}{\bibfnamefont{J.}~\bibnamefont{He}},
  \bibinfo{author}{\bibfnamefont{M.~S.} \bibnamefont{Hossain}},
  \bibinfo{author}{\bibfnamefont{X.}~\bibnamefont{Liu}}, \bibnamefont{et~al.},
  \bibinfo{journal}{Nature Materials} \textbf{\bibinfo{volume}{20}},
  \bibinfo{pages}{1353} (\bibinfo{year}{2021}), ISSN \bibinfo{issn}{1476-1122,
  1476-4660},
  \urlprefix\url{https://www.nature.com/articles/s41563-021-01034-y}.

\bibitem[{\citenamefont{Krix et~al.}(2022)\citenamefont{Krix, Scammell, and
  Sushkov}}]{krix_correlated_2022}
\bibinfo{author}{\bibfnamefont{Z.~E.} \bibnamefont{Krix}},
  \bibinfo{author}{\bibfnamefont{H.~D.} \bibnamefont{Scammell}},
  \bibnamefont{and} \bibinfo{author}{\bibfnamefont{O.~P.}
  \bibnamefont{Sushkov}}, \bibinfo{journal}{Physical Review B}
  \textbf{\bibinfo{volume}{105}}, \bibinfo{pages}{075120}
  (\bibinfo{year}{2022}), \bibinfo{note}{publisher: American Physical Society},
  \urlprefix\url{https://link.aps.org/doi/10.1103/PhysRevB.105.075120}.

\bibitem[{\citenamefont{Uri et~al.}(2020)\citenamefont{Uri, Grover, Cao,
  Crosse, Bagani, Rodan-Legrain, Myasoedov, Watanabe, Taniguchi, Moon
  et~al.}}]{uri_mapping_2020}
\bibinfo{author}{\bibfnamefont{A.}~\bibnamefont{Uri}},
  \bibinfo{author}{\bibfnamefont{S.}~\bibnamefont{Grover}},
  \bibinfo{author}{\bibfnamefont{Y.}~\bibnamefont{Cao}},
  \bibinfo{author}{\bibfnamefont{J.~A.} \bibnamefont{Crosse}},
  \bibinfo{author}{\bibfnamefont{K.}~\bibnamefont{Bagani}},
  \bibinfo{author}{\bibfnamefont{D.}~\bibnamefont{Rodan-Legrain}},
  \bibinfo{author}{\bibfnamefont{Y.}~\bibnamefont{Myasoedov}},
  \bibinfo{author}{\bibfnamefont{K.}~\bibnamefont{Watanabe}},
  \bibinfo{author}{\bibfnamefont{T.}~\bibnamefont{Taniguchi}},
  \bibinfo{author}{\bibfnamefont{P.}~\bibnamefont{Moon}}, \bibnamefont{et~al.},
  \bibinfo{journal}{Nature} \textbf{\bibinfo{volume}{581}}, \bibinfo{pages}{47}
  (\bibinfo{year}{2020}), ISSN \bibinfo{issn}{1476-4687},
  \urlprefix\url{https://www.nature.com/articles/s41586-020-2255-3}.

\bibitem[{\citenamefont{Wilson et~al.}(2020)\citenamefont{Wilson, Fu,
  Das~Sarma, and Pixley}}]{wilson_disorder_2020}
\bibinfo{author}{\bibfnamefont{J.~H.} \bibnamefont{Wilson}},
  \bibinfo{author}{\bibfnamefont{Y.}~\bibnamefont{Fu}},
  \bibinfo{author}{\bibfnamefont{S.}~\bibnamefont{Das~Sarma}},
  \bibnamefont{and} \bibinfo{author}{\bibfnamefont{J.~H.}
  \bibnamefont{Pixley}}, \bibinfo{journal}{Physical Review Research}
  \textbf{\bibinfo{volume}{2}}, \bibinfo{pages}{023325} (\bibinfo{year}{2020}),
  \urlprefix\url{https://link.aps.org/doi/10.1103/PhysRevResearch.2.023325}.

\bibitem[{\citenamefont{Polini et~al.}(2013)\citenamefont{Polini, Guinea,
  Lewenstein, Manoharan, and Pellegrini}}]{polini_artificial_2013}
\bibinfo{author}{\bibfnamefont{M.}~\bibnamefont{Polini}},
  \bibinfo{author}{\bibfnamefont{F.}~\bibnamefont{Guinea}},
  \bibinfo{author}{\bibfnamefont{M.}~\bibnamefont{Lewenstein}},
  \bibinfo{author}{\bibfnamefont{H.~C.} \bibnamefont{Manoharan}},
  \bibnamefont{and}
  \bibinfo{author}{\bibfnamefont{V.}~\bibnamefont{Pellegrini}},
  \bibinfo{journal}{Nature Nanotechnology} \textbf{\bibinfo{volume}{8}},
  \bibinfo{pages}{625} (\bibinfo{year}{2013}), ISSN \bibinfo{issn}{1748-3395},
  \urlprefix\url{https://www.nature.com/articles/nnano.2013.161}.

\bibitem[{\citenamefont{Huber et~al.}(2020)\citenamefont{Huber, Liu, Chen,
  Drienovsky, Sandner, Watanabe, Taniguchi, Richter, Weiss, and
  Eroms}}]{huber_gate-tunable_2020}
\bibinfo{author}{\bibfnamefont{R.}~\bibnamefont{Huber}},
  \bibinfo{author}{\bibfnamefont{M.-H.} \bibnamefont{Liu}},
  \bibinfo{author}{\bibfnamefont{S.-C.} \bibnamefont{Chen}},
  \bibinfo{author}{\bibfnamefont{M.}~\bibnamefont{Drienovsky}},
  \bibinfo{author}{\bibfnamefont{A.}~\bibnamefont{Sandner}},
  \bibinfo{author}{\bibfnamefont{K.}~\bibnamefont{Watanabe}},
  \bibinfo{author}{\bibfnamefont{T.}~\bibnamefont{Taniguchi}},
  \bibinfo{author}{\bibfnamefont{K.}~\bibnamefont{Richter}},
  \bibinfo{author}{\bibfnamefont{D.}~\bibnamefont{Weiss}}, \bibnamefont{and}
  \bibinfo{author}{\bibfnamefont{J.}~\bibnamefont{Eroms}},
  \bibinfo{journal}{Nano Letters} \textbf{\bibinfo{volume}{20}},
  \bibinfo{pages}{8046} (\bibinfo{year}{2020}), ISSN \bibinfo{issn}{1530-6984},
  \urlprefix\url{https://doi.org/10.1021/acs.nanolett.0c03021}.

\bibitem[{\citenamefont{Wang et~al.}(2020)\citenamefont{Wang, Reuter, Wieck,
  Hamilton, and Klochan}}]{wang_two-dimensional_2020}
\bibinfo{author}{\bibfnamefont{D.~Q.} \bibnamefont{Wang}},
  \bibinfo{author}{\bibfnamefont{D.}~\bibnamefont{Reuter}},
  \bibinfo{author}{\bibfnamefont{A.~D.} \bibnamefont{Wieck}},
  \bibinfo{author}{\bibfnamefont{A.~R.} \bibnamefont{Hamilton}},
  \bibnamefont{and} \bibinfo{author}{\bibfnamefont{O.}~\bibnamefont{Klochan}},
  \bibinfo{journal}{Applied Physics Letters} \textbf{\bibinfo{volume}{117}},
  \bibinfo{pages}{032102} (\bibinfo{year}{2020}), ISSN
  \bibinfo{issn}{0003-6951}, \bibinfo{note}{publisher: American Institute of
  Physics},
  \urlprefix\url{https://aip.scitation.org/doi/full/10.1063/5.0009462}.

\bibitem[{\citenamefont{Fürst et~al.}(2009)\citenamefont{Fürst, Pedersen,
  Flindt, Mortensen, Brandbyge, Pedersen, and Jauho}}]{furst_electronic_2009}
\bibinfo{author}{\bibfnamefont{J.~A.} \bibnamefont{Fürst}},
  \bibinfo{author}{\bibfnamefont{J.~G.} \bibnamefont{Pedersen}},
  \bibinfo{author}{\bibfnamefont{C.}~\bibnamefont{Flindt}},
  \bibinfo{author}{\bibfnamefont{N.~A.} \bibnamefont{Mortensen}},
  \bibinfo{author}{\bibfnamefont{M.}~\bibnamefont{Brandbyge}},
  \bibinfo{author}{\bibfnamefont{T.~G.} \bibnamefont{Pedersen}},
  \bibnamefont{and} \bibinfo{author}{\bibfnamefont{A.-P.} \bibnamefont{Jauho}},
  \bibinfo{journal}{New Journal of Physics} \textbf{\bibinfo{volume}{11}},
  \bibinfo{pages}{095020} (\bibinfo{year}{2009}), ISSN
  \bibinfo{issn}{1367-2630},
  \urlprefix\url{https://doi.org/10.1088/1367-2630/11/9/095020}.

\bibitem[{\citenamefont{Huber et~al.}(2021)\citenamefont{Huber, Steffen,
  Drienovsky, Sandner, Watanabe, Taniguchi, Pfannkuche, Weiss, and
  Eroms}}]{huber_brown-zak_2021}
\bibinfo{author}{\bibfnamefont{R.}~\bibnamefont{Huber}},
  \bibinfo{author}{\bibfnamefont{M.-N.} \bibnamefont{Steffen}},
  \bibinfo{author}{\bibfnamefont{M.}~\bibnamefont{Drienovsky}},
  \bibinfo{author}{\bibfnamefont{A.}~\bibnamefont{Sandner}},
  \bibinfo{author}{\bibfnamefont{K.}~\bibnamefont{Watanabe}},
  \bibinfo{author}{\bibfnamefont{T.}~\bibnamefont{Taniguchi}},
  \bibinfo{author}{\bibfnamefont{D.}~\bibnamefont{Pfannkuche}},
  \bibinfo{author}{\bibfnamefont{D.}~\bibnamefont{Weiss}}, \bibnamefont{and}
  \bibinfo{author}{\bibfnamefont{J.}~\bibnamefont{Eroms}},
  \bibinfo{journal}{arXiv:2106.11328 [cond-mat]}  (\bibinfo{year}{2021}),
  \urlprefix\url{http://arxiv.org/abs/2106.11328}.

\bibitem[{\citenamefont{Zhang et~al.}(2009)\citenamefont{Zhang, Tang, Girit,
  Hao, Martin, Zettl, Crommie, Shen, and Wang}}]{zhang_direct_2009}
\bibinfo{author}{\bibfnamefont{Y.}~\bibnamefont{Zhang}},
  \bibinfo{author}{\bibfnamefont{T.-T.} \bibnamefont{Tang}},
  \bibinfo{author}{\bibfnamefont{C.}~\bibnamefont{Girit}},
  \bibinfo{author}{\bibfnamefont{Z.}~\bibnamefont{Hao}},
  \bibinfo{author}{\bibfnamefont{M.~C.} \bibnamefont{Martin}},
  \bibinfo{author}{\bibfnamefont{A.}~\bibnamefont{Zettl}},
  \bibinfo{author}{\bibfnamefont{M.~F.} \bibnamefont{Crommie}},
  \bibinfo{author}{\bibfnamefont{Y.~R.} \bibnamefont{Shen}}, \bibnamefont{and}
  \bibinfo{author}{\bibfnamefont{F.}~\bibnamefont{Wang}},
  \bibinfo{journal}{Nature} \textbf{\bibinfo{volume}{459}},
  \bibinfo{pages}{820} (\bibinfo{year}{2009}), ISSN \bibinfo{issn}{1476-4687},
  \bibinfo{note}{number: 7248 Publisher: Nature Publishing Group},
  \urlprefix\url{https://www.nature.com/articles/nature08105}.

\bibitem[{\citenamefont{McCann and Koshino}(2013)}]{mccann_electronic_2013}
\bibinfo{author}{\bibfnamefont{E.}~\bibnamefont{McCann}} \bibnamefont{and}
  \bibinfo{author}{\bibfnamefont{M.}~\bibnamefont{Koshino}},
  \bibinfo{journal}{Reports on Progress in Physics}
  \textbf{\bibinfo{volume}{76}}, \bibinfo{pages}{056503}
  (\bibinfo{year}{2013}), ISSN \bibinfo{issn}{0034-4885},
  \urlprefix\url{https://doi.org/10.1088/0034-4885/76/5/056503}.

\bibitem[{\citenamefont{Hubbard and Flowers}(1964)}]{hubbard_electron_1964}
\bibinfo{author}{\bibfnamefont{J.}~\bibnamefont{Hubbard}} \bibnamefont{and}
  \bibinfo{author}{\bibfnamefont{B.~H.} \bibnamefont{Flowers}},
  \bibinfo{journal}{Proceedings of the Royal Society of London. Series A.
  Mathematical and Physical Sciences} \textbf{\bibinfo{volume}{281}},
  \bibinfo{pages}{401} (\bibinfo{year}{1964}), \bibinfo{note}{publisher: Royal
  Society},
  \urlprefix\url{https://royalsocietypublishing.org/doi/10.1098/rspa.1964.0190}.

\bibitem[{\citenamefont{Keser et~al.}(2022)\citenamefont{Keser, Lyanda-Geller,
  and Sushkov}}]{keser_nonlinear_2022}
\bibinfo{author}{\bibfnamefont{A.~C.} \bibnamefont{Keser}},
  \bibinfo{author}{\bibfnamefont{Y.}~\bibnamefont{Lyanda-Geller}},
  \bibnamefont{and} \bibinfo{author}{\bibfnamefont{O.~P.}
  \bibnamefont{Sushkov}}, \bibinfo{journal}{Physical Review Letters}
  \textbf{\bibinfo{volume}{128}}, \bibinfo{pages}{066402}
  (\bibinfo{year}{2022}), \bibinfo{note}{publisher: American Physical Society},
  \urlprefix\url{https://link.aps.org/doi/10.1103/PhysRevLett.128.066402}.

\bibitem[{\citenamefont{Hwang and Das~Sarma}(2008)}]{hwang_screening_2008}
\bibinfo{author}{\bibfnamefont{E.~H.} \bibnamefont{Hwang}} \bibnamefont{and}
  \bibinfo{author}{\bibfnamefont{S.}~\bibnamefont{Das~Sarma}},
  \bibinfo{journal}{Physical Review Letters} \textbf{\bibinfo{volume}{101}},
  \bibinfo{pages}{156802} (\bibinfo{year}{2008}), \bibinfo{note}{publisher:
  American Physical Society},
  \urlprefix\url{https://link.aps.org/doi/10.1103/PhysRevLett.101.156802}.

\bibitem[{\citenamefont{Dean et~al.}(2010)\citenamefont{Dean, Young, Meric,
  Lee, Wang, Sorgenfrei, Watanabe, Taniguchi, Kim, Shepard
  et~al.}}]{dean_boron_2010}
\bibinfo{author}{\bibfnamefont{C.~R.} \bibnamefont{Dean}},
  \bibinfo{author}{\bibfnamefont{A.~F.} \bibnamefont{Young}},
  \bibinfo{author}{\bibfnamefont{I.}~\bibnamefont{Meric}},
  \bibinfo{author}{\bibfnamefont{C.}~\bibnamefont{Lee}},
  \bibinfo{author}{\bibfnamefont{L.}~\bibnamefont{Wang}},
  \bibinfo{author}{\bibfnamefont{S.}~\bibnamefont{Sorgenfrei}},
  \bibinfo{author}{\bibfnamefont{K.}~\bibnamefont{Watanabe}},
  \bibinfo{author}{\bibfnamefont{T.}~\bibnamefont{Taniguchi}},
  \bibinfo{author}{\bibfnamefont{P.}~\bibnamefont{Kim}},
  \bibinfo{author}{\bibfnamefont{K.~L.} \bibnamefont{Shepard}},
  \bibnamefont{et~al.}, \bibinfo{journal}{Nature Nanotechnology}
  \textbf{\bibinfo{volume}{5}}, \bibinfo{pages}{722} (\bibinfo{year}{2010}),
  ISSN \bibinfo{issn}{1748-3395}, \bibinfo{note}{number: 10 Publisher: Nature
  Publishing Group},
  \urlprefix\url{https://www.nature.com/articles/nnano.2010.172}.

\end{thebibliography}
\end{document}